\documentclass[12pt,journal,final,letterpaper,onecolumn]{article}
\usepackage[top=0.86in, bottom=0.86in, left=0.88in, right=0.88in]{geometry}
\usepackage{graphicx} % more modern
\usepackage{subfigure}
\usepackage{algorithm}
\usepackage{algorithmic}
\usepackage{hyperref}
\usepackage{amsmath,  amssymb, amsthm}
\usepackage[comma,authoryear]{natbib}

\usepackage{color}

\usepackage{setspace}
\newcommand{\beq}{\vspace{0mm}\begin{equation}}
\newcommand{\eeq}{\vspace{0mm}\end{equation}}
\newcommand{\beqs}{\vspace{0mm}\begin{eqnarray}}
\newcommand{\eeqs}{\vspace{0mm}\end{eqnarray}}
\newcommand{\barr}{\begin{array}}
\newcommand{\earr}{\end{array}}

\newcommand{\Imat}{{\bf I}}

\newcommand{\xv}{\boldsymbol{x}}

\newcommand{\zv}{\boldsymbol{z}}

\newcommand{\muv}[0]{{\boldsymbol{\mu}}}

\newcommand{\E}{\mathbb{E}}

\newtheorem{thm}{Theorem} %[section]
\newtheorem{cor}[thm]{Corollary}

\usepackage{natbib,natbibspacing}
\begin{document}
\title{Generalized Negative Binomial Processes and\\ the Representation of Cluster Structures} \author{Mingyuan~Zhou\thanks{\emph{Address for correspondence}: Department of Information, Risk, and Operations Management, 2110 Speedway Stop B6500, Austin, TX 78712, USA.
\emph{Email:} \texttt{mingyuan.zhou@mccombs.utexas.edu}
% The author's work was supported by the Department of Information, Risk, and Operations Management, McCombs School of Business and the Division of Statistics and Scientific Computation at the University of Texas at Austin.}
}\\ 
%Department of Information, Risk, and Operations Management\\
McCombs School of Business, The University of Texas at Austin%, Austin, TX 78712, USA\\
%\texttt{mingyuan.zhou@mccombs.utexas.edu} \\%\\
}

\maketitle

\vspace{-3mm}
\begin{abstract}
The paper introduces the concept of a cluster structure to define a joint distribution of the sample size and its exchangeable random partitions. 
The cluster structure 
allows the probability distribution of the random partitions of a subset of the sample to 
be dependent on the sample size, a feature not presented in   a partition structure. A generalized negative binomial process count-mixture model is proposed to generate a cluster structure, where in the prior the number of clusters is finite and Poisson distributed and the cluster sizes 
follow a truncated negative binomial distribution. 
The number and  sizes of clusters 
can be controlled to exhibit  distinct asymptotic behaviors. Unique model properties 
are illustrated with example  
clustering results using a generalized P\'{o}lya  urn sampling scheme. 
The paper provides   new methods to generate exchangeable random partitions 
and to control both the cluster-number and cluster-size distributions. 

\emph{Keywords}: %Asymptotics, 
Bayesian nonparametrics, %completely random measures,  
%cluster structure, %clustering, %
%consistency, 
clustering,
count-mixture modeling, %Dirichlet process, 
%exchangeable random partitions,  
 exchangeable cluster/partition probability functions, generalized Chinese restaurant/gamma/negative binomial processes, partition structure. %, power-law %, generalized negative binomial process. %gamma process, generalized Chinese restaurant process, generalized gamma process, generalized negative binomial process,  mixture modeling,  normalized random measures, truncated negative binomial distribution.\end{keywords}

\end{abstract}

%\vspace{-6mm}
\section{Introduction}
A foundation of contemporary
probabilistic  clustering algorithms is sampling consistency, which requires the  probability distribution of the random partitions of a subset of size $j$ to be the same as that of the random partitions of $j$ elements, uniformly at random subsampled without replacement from a sample of size $m\ge j$. In other words, in the prior, the clustering of a subset with $j$ elements is independent of the number of elements remaining in the data set. In practice, however, it is possible that the clustering  is directly related to the sample size. E.g., to achieve an 8\% projected annual return with minimized risk by investing in Lending Club, an online peer-to-peer lending platform where each note is purchased at 25 dollars,   the partition of one thousand dollar total investment into 40 notes and the partition of one thousand dollars from a total investment of one million dollars could be drastically different.
%real examples where 
It is also common in real applications that 
the clusters of the current sample may be deleted, split or merged, as additional data points are observed. E.g., the Olinguito, which was originally classified as the Olingos, has been recently identified as a new mammal species and has been further classified into four subspecies \citep{helgen2013taxonomic}. Relaxing the constraint of sampling consistency, the paper aims to build a clustering strategy that takes the sample size into careful consideration.

Motivated by the success of the Ewens sampling formula \citep{ewens1972sampling} in population genetics, where a sample is subsampled from a large population, \citet{kingman1978random,kingman1978representation} introduced the concept of a partition structure.  Defining a family of %sequence of 
consistent probability distributions for random partitions of positive integers, %  that are consistent as the sample size $m$ varies, 
the partition structure is commonly featured in %common feature of existing %nonparametric Bayesian 
probabilistic clustering algorithms.  
%With the random probability measure marginalized out, the ties between the elements of a sample of size $m$ define 
For a random partition $\Pi_m=\{A_1,\cdots,A_l\}$ of the set $[m]:=\{1,\cdots,m\}$, where $i\in A_k$ indicates that element $i$ belongs to cluster $k=1,\cdots,l$,
the consistency of the probability functions $P(\Pi_1),\cdots,P(\Pi_\infty)$ 
%of $P(\Pi_1),\cdots,P(\Pi_\infty)$ 
requires that for a subset of size $j$ in a sample of size $m\ge j$, its partition probability function $P(\Pi_{j}|m)$, which governs %, which is 
the probability distribution of the random partitions $(\Pi_j|m)$ obtained by marginalizing out $m-j$ elements %uniformly at random %$m$ 
from $\Pi_m$, is the same for any $m\in\{j,j+1,\cdots\}$. Thus in a partition structure,  %the partition probability function 
$P(\Pi_j|m)$ is equal to $P(\Pi_j)$ 
 and  is independent of the sample size $m$. 
 As further developed in \citep{pitman1995exchangeable,csp}, if %the probability for an arbitrary partition $\Pi_m$ 
 $P(\Pi_m)$ depends only on the number and sizes of $A_k$, regardless of their orders, then it is called an exchangeable partition probability function (EPPF) of $\Pi_m$, expressed as  $P(\Pi_m=\{A_1,\cdots,A_l\})=p_m(n_1,\cdots,n_l)$, where $n_k=|A_k|$. 
The sampling consistency  %of $P(\Pi_m)$ as $m$ varies %
amounts to an addition rule \citep{csp,Gnedin_deletion} for %
the EPPF $P(\Pi_m)$ %is required %of a partition structure 
%to satisfy $p_1(1) = 1$ and an addition rule as
that $p_1(1) = 1$ and
\beqs\label{eq:addrule}
p_m(n_1,\cdots,n_l) =  p_{m+1}(n_1,\cdots,n_l,1)+ \sum_{k=1}^l p_{m+1}(n_1,\cdots,n_k+1,\cdots,n_l),
\eeqs
with which $\Pi_{m+1}$ can be constructed from  $\Pi_m$ by assigning element $(m+1)$ to $A_{z_{m+1}}$ based on the prediction rule as
$
z_{m+1}|\Pi_m \sim \frac{p_{m+1}(n_1,\cdots,n_l,1)}{p_m(n_1,\cdots,n_l)}\delta_{l+1} +  \sum_{k=1}^l \frac{p_{m+1}(n_1,\cdots,n_k+1,\cdots,n_l)}{p_m(n_1,\cdots,n_l)}\delta_{k}.
$ This paper calls an EPPF $P(\Pi_m)$  that satisfies (violates) the addition rule as a consistent (inconsistent) EPPF, which is referred as an infinite (finite) EPPF in \citet{csp}.

Moving beyond the Ewens sampling formula,  
various  approaches, including the Pitman-Yor process 
 \citep{perman1992size,pitman1997two}, Poisson-Kingman models \citep{pitman2003poisson}, species sampling \citep{Pitman96somedevelopments,lee2013},  stick-braking priors \citep{ishwaran2001gibbs}, and Gibbs-type random partitions \citep{gnedin2006exchangeable},   have been proposed to construct more general partition structures.  
See \citet{muller2004nonparametric}, \citet{BeyondDP} and \citet{Muller2013} for reviews. 
Among these approaches,
there has been increasing interest in 
normalized random measures with independent increments (NRMIs)
\citep{regazzini2003distributional,lijoi2005inverseGaussian,lijoi2007controlling,james2009posterior}, 
where a completely random measure with a finite and strictly positive total random mass is normalized to construct a random probability measure. For example, the normalized gamma process is a Dirichlet process \citep{ferguson73}, the marginalization of which leads to a variant of the Ewens sampling formula \citep{DP_Mixture_Antoniak}. More advanced completely random measures, such as the generalized gamma process of \citet{brix1999generalized}, can be employed to produce more general consistent exchangeable random partitions \citep{pitman2003poisson,csp,lijoi2007controlling}, but the expressions of the consistent EPPF and its associated prediction rule usually involve integrations that are difficult to calculate. In addition, the construction of the random probability measure is independent of the sample size and due to normalization,  the scale and mass parameters of the completely random measure become redundant to each other and the total random mass has to be strictly positive. % 

Recognizing the need to revise the probability distribution of the random partitions of a subset of size $j$ along with the change of the sample size $m$, %as the sample size  $m$ increases, 
we propose a novel nonparametric Bayesian approach that allows
%To make 
 the 
partition probability function $P(\Pi_j|m)$ % of a subset of the sample 
to be dependent on $m$. If $P(\Pi_j|m)$ is exchangeable in its arguments, we refer it as a size-dependent EPPF.
We model the sample size $m$ as a Poisson random variable, %from a Poisson distribution, 
whose mean is parameterized by the total random mass of a completely random measure $G$ over a measurable space $\Omega$. The total random mass $G(\Omega)$ %is finite almost surely (a.s.) and 
is used to normalize $G$ to obtain a random probability measure $G/G(\Omega)$, based on which the $m$  points are then clustered. 
Linking $m$ to $G(\Omega)$ with a Poisson distribution not only introduces dependence between the normalized random probability measure and the sample size, %be dependent on the sample size, 
but also makes the scale of $G$ become identifiable.
As further discussed, the new model requires  $G(\Omega)$ to be finite but not necessarily strictly positive.  
With $G$ marginalized out, the joint distribution of  $m$ and its exchangeable random partition $\Pi_m$ is called an exchangeable cluster probability function (ECPF). %, %, which is in a fully factorized form 
On observing a sample of size $m$, the EPPF $P(\Pi_m)$ directly comes from the ECPF $P(\Pi_m,m)$ divided by the marginal distribution of $m$, and the size-dependent EPPF $P(\Pi_j|m)$ is derived by marginalizing $m-j$ elements from $P(\Pi_m)$.

Distinct from a partition structure that requires $P(\Pi_j|m)\equiv P(\Pi_j)$ and hence a consistent EPPF $P(\Pi_m)$ that satisfies the addition rule, the introduced model allows $P(\Pi_j|m)$ to be dependent on the sample size $m$ and permits the generated exchangeable random partitions $\Pi_m$ to be inconsistent in distribution as $m$ varies.
We call the introduced model % constructed in this way 
as a count-mixture model, since as further shown in the paper, 
the \emph{a priori} number of data points $X(A)$ on each measurable subset $A\subset \Omega$ follows a Poisson distribution parameterized by $G(A)$, and both a Poisson mixture and a compound Poisson distributions if $G(A)$ is marginalized out.
%allow the total random mass of the completely random measure to have a nonzero probability to be zero, %exchangeable random partitions, %a sample of arbitrary size, 
 We introduce the concept of a cluster structure to characterize the count-mixture model, which    extends Kingman's concept of a partition structure to allow $P(\Pi_j|m)\neq P(\Pi_j)$ for $m>j$. A cluster structure specifies in its prior that 
%in the prior 
the number of clusters follows a Poisson distribution with a finite mean, 
the sizes of these clusters are independently and identically (iid) drawn from a positive discrete distribution, and hence the sample size 
$m$ follows a compound Poisson distribution. On observing a set of $j$ data points,  the model provides a size-dependent EPPF $P(\Pi_j|m)$ that can be modified  to reflect the variation of the sample size $m=j,j+1,\cdots$. 

We consider a generalized negative binomial (NB) process count-mixture model where
$G$ is drawn from a generalized gamma process \citep{brix1999generalized}. 
 A draw from the generalized NB process (gNBP) represents a cluster structure with a Poisson distributed finite number of clusters, whose sizes 
follow a truncated NB distribution. Marginally the sample size 
follows a generalized NB distribution.  The EPPF of the gNBP generally violates the addition rule. 
A stochastic process with this EPPF is referred as the generalized Chinese restaurant process (CRP), which has  a simple prediction rule that is used to develop a generalized  P\'{o}lya urn sampling scheme. 
These three count distributions and the prediction rule are %fully specified and explicitly controlled
%fully
determined by a discount, a probability and a mass parameters. %, which can all be conveniently inferred from the data. % without numerical stability issues.
%Among them, %these three parameters,
The discount parameter %plays a critical role in controlling %both the tail behavior of the distribution of %the number of samples on each distinct atom
%cluster sizes and
controls the asymptotic behaviors of both the number and sizes of clusters, whereas how the mass parameter is modeled determines whether the asymptotic behavior of the cluster number is similar or completely opposite to that of the normalized generalized gamma process mixtures in \citet{lijoi2007controlling,james2009posterior}.

The remainder of the paper is organized as follows. Section \ref{sec:preliminary} briefly reviews some background materials. 
Section \ref{sec:CountMixture} introduces the concept of a cluster structure and constructs the count-mixture modeling framework.  
Section \ref{sec:gNBP}  introduces the  gNBP, discusses how to control its asymptotic behaviors, and shows how its partition probability functions are dependent on the sample size. Section \ref{sec:inference} presents 
a generalized P\'{o}lya urn sampling scheme  
for posterior simulation %algorithms 
and Section \ref{sec:Results} presents example clustering results.

\section{Preliminaries}\label{sec:preliminary}

\subsection{Generalized Gamma Process}

A completely random measure \citep{Kingman,PoissonP,Wolp:Clyd:Tu:2011} $G$ defined on the product space $\mathbb{R}_+\times \Omega$, where $\mathbb{R}_+=\{x:x\ge0\}$ and $\Omega$ is a measurable space, assigns independent infinitely divisible random variables $G(A_i)$ to disjoint Borel sets $A_i\subset\Omega$, with characteristic functions
$ %\beqs\label{LevyCF}
\E\big[e^{iuG(A)}\big]=e^{\int\hspace{-1mm}\int_{\mathbb{R}_+\times A}(e^{iur}-1)\nu(dr d\omega)}.
$ %\eeqs
%where $\nu(dr d\omega)\equiv \nu^+ \pi(dr d\omega)$. 
%More genera lly, 
If the L\'{e}vy measure $\nu({dr d\omega})$  satisfies %the local $L_1$ integrability condition
$ %\beqs \label{eq:LevyL1Condition}
\int\hspace{-2mm}\int_{\mathbb{R}_+\times A}\min\{1, |r|\}\nu(dr d\omega) < \infty
$ %\eeqs
for any %measurable subset 
$A\subset \Omega$, then $G$ is well defined, even if the Poisson intensity $\nu^+=\nu(\mathbb{R}_+\times \Omega)$ is infinite. For simplicity, we consider $G$ to be a homogenous completely random measure, whose atoms' weights are independent to their locations.  We write the L\'{e}vy measure of a homogenous $G$ as 
   $
   \nu(drd\omega)=\rho(dr)G_0(d\omega),\notag
   $
    where $G_0$ is a continuous base measure over $\Omega$, with a finite total mass $\gamma_0=G_0(\Omega)$.

The gamma process \citep{ferguson73,Wolpert98poisson/gammarandom}  and beta process \citep{Hjort,JordanBP,%Mingyuan09,
dHBP_AISTATS2011,BPFA_TIP2012} are two commonly used completely random measures. The widely used Dirichlet process \citep{ferguson73,DP_Mixture_Antoniak,Escobar1995} is a normalized gamma process. 
Another example is the generalized gamma process  $G\sim\mbox{gGaP}(G_0,a,c)$ %is a completely random measure 
defined on the product space $\mathbb{R}_+\times \Omega$, where $a< 1$ is a discount parameter and $1/c$ is a scale parameter \citep{brix1999generalized}. %, and $G_0$ is a finite and continuous base measure over $\Omega$ .  
It assigns independent infinitely divisible generalized gamma random variables $G(A_j)\sim{{}}\mbox{gGamma}(G_0(A_j),a,1/c)$ to disjoint Borel sets $A_j\subset \Omega$, %. For each subset $A\subset\Omega$, $G(A)$ follows a generalized gamma distribution,
with Laplace transforms
\beq\label{eq:Laplace}
%L_{a,c,G_0(A)}(s)
\E[e^{-sG(A)}] = e^{-\frac{G_0(A)}{a}\left[(c+s)^a-c^a\right]}.
\eeq
The generalized gamma distribution $x\sim{{}}\mbox{gGamma}(\gamma,a,1/c)$, with $\E[x]=\gamma c^{a-1}$, 
 was independently suggested by \citet{tweedie1984index} and \citet{hougaard1986survival} and also studied in \citet{bar1986reproducibility,alen1992modelling,jorgensen1997theory}.
As $a\rightarrow 0$, since $\lim_{a\rightarrow 0}\frac{1-(1-p)^a}{ap^a} = -\ln(1-p)$ and hence $ %\beq
%L_{a,c,G_0(A)}(s)
\lim_{a\rightarrow 0}\E[e^{-sx}] = (1+ s/c)^{-\gamma}
$, the generalized gamma distribution  becomes a gamma distribution  $x\sim\mbox{Gamma}(\gamma,1/c)$, with $\E[x]=\gamma/c$. A generalized gamma distribution scaled with a positive constant $\beta>0$ is distributed as $\beta x\sim\mbox{gGamma}(\gamma\beta^a,a,\beta/c)$.

Using (\ref{eq:Laplace}), %(\ref{LevyCF}),
the L\'{e}vy measure of the generalized gamma process
%$G\sim{{}}\mbox{gGaP}(a,1/c,G_0)$
can be expressed as
\beqs\label{eq:LevyGGP}
\nu(dr d\omega) = \frac{1}{\Gamma(1-a)}r^{-a-1}e^{-cr}dr G_0(d\omega).
\eeqs
%where %$a\in[0,1)$ is a discount parameter, $c$ is the concentration parameter,
%$\Gamma(\cdot)$ is the gamma function;
When $a\rightarrow0$, we recover the gamma process; if $0<a<1$ and $c\rightarrow 0$, we recover the $a$-stable process \citep{pitman2003poisson,lijoi2007controlling}; and if $a=1/2$, we recover the inverse Gaussian process \citep{lijoi2005inverseGaussian}.  A draw from %the generalized  gamma process
$G\sim{{}}\mbox{gGaP}(G_0,a,c)$ %consists of countably infinite points,
 can be expressed as
\beq
G = \sum_{k=1}^{K} r_k \delta_{\omega_k},K\sim\mbox{Pois}(\nu^+),~(r_k,\omega_k)\stackrel{iid}{\sim} \pi(drd\omega),\notag
\eeq
where $r_k=G(\omega_k)$ is the weight for atom $\omega_k$ %, $G(\Omega)=\sum_{k=1}^{K} r_k$ is the total random mass, 
and $\pi(dr d\omega)\nu^{+} \equiv \nu(drd\omega)$. 
Except where otherwise specified, 
we consider $a<1$ and $c>0$ %or $0<a<1$ and $c=0$
 in this paper.
If $0\le a<1$, since the Poisson intensity $\nu^+ = \nu(\mathbb{R}_+\times \Omega) = \infty$ ($i.e.$, $K=\infty$) and
  $
 \int\hspace{-2mm}\int_{\mathbb{R}_+\times \Omega} \min\{1, r\} \nu(dr d\omega)  %= \gamma_0c^{a-1}
 $
 is finite, a drawn from ${{}}\mbox{gGaP}(G_0,a,c)$ consists of countably infinite atoms. On the other hand, if $a<0$, then $\nu^+=\frac{\gamma_0c^a}{-a}$ and thus $K\sim \mbox{Pois}(\frac{\gamma_0c^a}{-a})$ ($i.e.$, $K$ is finite a.s.) and $r_k\stackrel{iid}{\sim}\mbox{Gamma}(-a,1/c)$.

\subsection{Normalized Random Measure %with Independent Increments
Mixtures}\label{NRMI}

An NRMI mixture model \citep{regazzini2003distributional,BeyondDP} emploies
%Since 
a normalized completely random measure $\widetilde{G}=G/G(\Omega)$ %is a random probability measure, $\widetilde{G}$ can be employed %for mixture modeling, where  %$G\sim{{}}\mbox{gGaP}(a,c,G_0)$
to model the density of a data point $x$ as
$ %\beqs\label{eq:fxG}
%&
f(x|G) = \int_{\Omega} \kappa(x|\omega)d \widetilde{G}(\omega) = \sum_{k=1}^{K} \frac{r_k}{G(\Omega)} \kappa(x|\omega_k),\notag
$ % \eeqs
where 
$\kappa(x|\omega)$ is a  density function  for $x$  with parameter   $\omega$, $G(\Omega)=\sum_{k=1}^{K} r_{k}$ is the total random mass that is required to be finite and strictly positive, and $K={\infty}$ is the total number of atoms. 
Note that the strict positivity of $G(\Omega)$ implies that $\nu^+=\infty$ and hence $K=\infty$ \citep{regazzini2003distributional,BeyondDP}. By introducing a categorical latent  variable 
$
z|G \sim \sum_{k=1}^{{K}}\frac{r_k}{G(\Omega)}\delta_k, \notag %~z=1,\cdots,K,
$ 
 we can augument %(\ref{eq:fxG})
$f(x|G)$ as
$
f(x,z|G) =f(x|z,G)f(z|G) = \kappa(x|\omega_{z}) \frac{r_{z}}{\sum_{k=1}^{K} r_{k}}.
$
When a sample of $m$ data points  %$\xv=(x_1,\cdots,x_m)$
$\{x_i\}_{1,m}$ is observed, we have
%\begin{align}
\beqs\label{eq:f_G_N}
f(\zv|G,m)= \prod_{i=1}^m \frac{r_{z_i}}{\sum_{k=1}^{K} r_k}
=  \frac{1}{(\sum_{k=1}^{K} r_k)^{m}}%\prod_{k=1}^{K} n_k!}
\prod_{k=1}^{K} {r_k^{n_k}} , %\\
%&= \int_0^{K} \frac{1}{\Gamma(N)}v^{N-1}e^{-v\sum_{k=1}^{K} r_k}dv \prod_{k=1}^{K} r_k^{n_k} \prod_{i=1}^N \kappa(x_i|\omega_{z_i})
%\end{align}
\eeqs
where  $\zv=(z_1,\cdots,z_m)$ is a sequence of categorical random variables  indicating the cluster memberships of data points $\xv=(x_1,\cdots,x_m)$, $n_k = \sum_{i=1}^m \delta (z_{i}=k)$ is the number of data points assigned to $\omega_k$ and $m=\sum_{k=1}^{K} n_k$. The ties between $z_i$ define a random partition $\Pi_m$ of $[m]$.

Posterior simulation of $G$ based on (\ref{eq:f_G_N}) %either (\ref{eq:f_G_N}) or (\ref{eq:f_G_N1})
 is usually challenging as the term ${(\sum_{k=1}^{K} r_k)^{-1}}$ is not in a factorized form and the scale of $G$ lacks identifiability. % can be freely scaled and. %, especially if $K=\infty$. %To make Markov chain Monte Carlo (MCMC) inference more amenable for (\ref{eq:f_G_N}),
 Following \citet{%nieto2004normalized,
james2009posterior}, %,griffin2011posterior%,favaromcmc
 a specific auxiliary variable  $\beta>0$, with %(v|m,\sum_{k=1}^{K} r_k)
 $ %\beq\label{eq:auxiliary}
(\beta|m,G(\Omega)) \sim\mbox{Gamma}(m,1/G(\Omega))
$ %\eeq 
and thus $\E[\beta|m,G(\Omega)] ={m}/{G(\Omega)}$, %\sum_{k=1}^{K} r_k)$,
can be introduced to yield a %fully factorized likelihood as
likelihood as
%\begin{align}
\beqs\label{eq:fxzv_G_N}
f(\zv,\beta|G,m)%&= %\prod_{i=1}^N \frac{r_{z_i}}{\sum_{k=1}^{K} r_k} \kappa(x_i|\omega_{z_i})
%= \Big(\sum_{k=1}^{K} r_k\Big)^{-N} \prod_{k=1}^{K} r_k^{n_k} \prod_{i=1}^N \kappa(x_i|\omega_{z_i}) \\
= %\int_0^{K}
 \frac{\beta^{m-1}}{(m-1)!}e^{-\beta\sum_{k=1}^{K} r_k} \prod_{k=1}^{K} r_k^{n_k}.
 %\frac{v^{N-1}}{\Gamma(n)}e^{-v\sum_{k=1}^{K} r_k} \prod_{k=1}^{K} {r_k^{n_k}} \prod_{i=1}^N \kappa(x_i|\omega_{z_i}).
%\end{align}
\eeqs
Marginalizing out $G$ leads to
\beq\label{eq:betaEPPF}
f(\zv,\beta|m,\gamma_0,\rho) = 
 \frac{\beta^{m-1}}{(m-1)!}  e^{\gamma_0\int_0^\infty (e^{-\beta r}-1)\rho(dr)} \prod_{k:\omega_k\in\mathcal{D}_m} \gamma_0 \int_0^\infty r^{n_k} e^{-\beta r} \rho(dr),
\eeq
where $\mathcal{D}_m=\{\omega_k\}_{k:n_k>0}$ includes all the points of discontinuity occupied by the $m$ data points. %This EPPF is consistent as $m$ varies \citep{csp,lijoi2007controlling}. 
%As in \citet{pitman2003poisson,csp,lijoi2007controlling,james2009posterior}, 
Further marginalizing out $\beta$ yields a consistent EPPF \citep{pitman2003poisson,csp}, expressed  as
$ %\beq\label{eq:consistentEPPF}
f(\zv|m,\gamma_0,\rho) = \int_0^\infty f(\zv,\beta|m,\gamma_0,\rho) d\beta,
%\int_0^\infty \frac{\beta^{m-1}}{(m-1)!}  e^{\gamma_0\int_0^\infty (e^{-\beta r}-1)\rho(dr)} \prod_{k:\omega_k\in\mathcal{D}_m} \gamma_0 \int_0^\infty r_k^{n_k} e^{-\beta r_k} \rho(dr_k)\  d\beta,
$ where the integral %in (\ref{eq:consistentEPPF}) 
is usually not analytic. 
For $G\sim\mbox{gGaP}(a,c,G_0)$, 
\citet{lijoi2007controlling} derived the analytic expressions of the EPPF % of (\ref{eq:consistentEPPF}) 
and the associated prediction rule. 
These analytical expressions, however, are still not easy to calculate.
To simplify the computation, by conditioning on the auxiliary variable $\beta$, %with the assistance of the auxiliary variable $\beta$, %by using  the auxiliary variable  $(\beta|m,G(\Omega)) \sim\mbox{Gamma}(m,1/G(\Omega))$, % in (\ref{eq:auxiliary}), 
\citet{james2009posterior} further developed a generalized Blackwell-MacQueen sampling scheme by exploiting the fully factorized form of %(\ref{eq:fxzv_G_N}) and 
(\ref{eq:betaEPPF}). Conditioning on the auxiliary variable, the EPPF of an NRMI mixture model is usually inconsistent but often becomes more amenable to posterior simulation  \citep{james2009posterior}, stimulating the development of a number of posterior simulation algorithms including % have been constructed based on (\ref{eq:fxzv_G_N}) %the likelihood shown above %in (\ref{eq:fxzv_G_N})
\citet{griffin2011posterior,barrios2012modeling,%griffin2011sequential,
 favaromcmc}.  For the proposed count-mixture models, we show below that inconsistent %and size-dependent 
 exchangeable random partitions can be constructed without the need of an auxiliary variable.

\section{Count-Mixture Modeling and Cluster Structures}\label{sec:CountMixture}

Distinct from a partition structure, 
a cluster structure defines a probability distribution of all possible exchangeable random partitions, which are not constrained to be consistent in distribution as the sample size varies. %The sampling consistency %between different sample sizes 
%is desired if a sample of size $m$ is randomly subsampled from a large population. 
Without imposing sampling consistency, there is a need to define a new mechanism to generate a sample of arbitrary size; and as discussed in Section~\ref{NRMI}, the completely random measure $G$  in an NRMI mixture model is independent of the sample size and %as its normalization is scale invariant and 
$G(\Omega)$ has to be strictly positive. % can be freely scaled. % with a positive constant $\beta$. 
To resolve these issues, we propose a ``two birds with one stone'' solution where we link the sample size $m$ to the total random mass of $G$ with a Poisson distribution as $m\sim\mbox{Pois}(\beta G(\Omega))$, \emph{i.e.}, $\E[m|\beta,G(\Omega)]= \beta G(\Omega)$, where $\beta$ is a positive scale parameter.
Since the $m$ data points are further clustered based on the normalized random probability measure $G/G(\Omega)=\sum_{k=1}^K \frac{r_k}{\sum_{k=1}^K r_k}\delta_{\omega_k}$, %using Lemma 4.1 of \citet{BNBP_PFA_AISTATS2012}, 
equivalently we have 
\beq
m=\sum_{k=1}^K n_k,~n_k\sim\mbox{Pois}(\beta r_k).
\eeq Therefore, letting $m\sim\mbox{Pois}(\beta G(\Omega))$ directly links the cluster sizes $n_k$ to weights $r_k$ with  Poisson distributions \citep{BNBP_PFA_AISTATS2012}. The mechanism to generate a sample of arbitrary size is now well defined and $G$ can no longer be scaled freely.  The new construction also allows $G(\Omega)=0$, for which $m\equiv 0$ and there is no more need to calculate $G/G(\Omega)$.  Allowing $G(\Omega)=0$ with a nonzero probability relaxes the requirement of $\nu^+=\infty$ (\emph{i.e.}, $K=\infty$). %  is no longer mandatory.  
We call a model constructed in this way as a count-mixture model.

\subsection{Count-Mixture Modeling}
We construct a cluster structure via a count-mixture model as 
\beq\label{eq:CRMcountandmixture}
x_i\sim\kappa(\omega_{z_i}),~\omega_k\sim g_0,~ z_i\sim\sum_{k=1}^K \frac{r_{k}}{G(\Omega)} \delta_k, %f(z_i|G)={r_{z_i}}/{G(\Omega)},
~m\sim{\mbox{Pois}}(\beta G(\Omega)),
 \eeq
 where $i=1,\cdots,m$, 
  $g_0(d\omega):=G_0(d\omega)/\gamma_0$ is the base distribution and $K\le \infty$.  Using (\ref{eq:f_G_N}), the joint conditional likelihood of $\zv=(z_1,\cdots,z_m)$ and $m$ becomes
\begin{align}\label{eq:f_G6}
f(\zv,m|\beta, G) &= f(\zv|G,m)\mbox{Pois}(m; \beta G(\Omega)) = \frac{\beta^{m}}{m!}e^{-\beta \sum_{k=1}^K r_k} \prod_{k=1}^{K} {r_k^{n_k}}
\end{align}
and hence the conditional likelihood of $n_{1:K}=(n_1,\cdots,n_K)$ and $m$ can be expressed as
\beq\label{eq:f_G5}
f(n_{1:K},m|\beta,G) = \frac{m!}{\prod_{k=1}^{K} n_k!}f(\zv,m|\beta,G) =\delta_{\sum_{k=1}^K n_k}(m)\prod_{k=1}^K \mbox{Pois}(n_k;\beta r_k),
\eeq
where $\delta_n(m)$ is a unit point mass at $m=n$.
The conditional likelihoods in  (\ref{eq:f_G6}) and (\ref{eq:f_G5}) are fully factorized by construction, % in a count-mixture model, 
a highly desired property for posterior simulation. Distinct from \citet{james2009posterior}, %a count-mixture model does not need auxiliary variables and allows $K<\infty$. %there is no need to introduce 
here an auxiliary variable is not needed and $K<\infty$ is permitted.

 Note that $\beta$ in  (\ref{eq:fxzv_G_N}) is an auxiliary variable that has to be sampled or marginalized out % using (\ref{eq:auxiliary}) or $\E_G[f(\zv,\beta|G,m)]$ 
and hence shall never be fixed. Whereas  $\beta$ in (\ref{eq:f_G6}) is a scale parameter that can be absorbed into   the parameters of $G$ and hence can be fixed at any positive constants.
Without loss of generality, we  first fix $\beta\equiv 1$ in count-mixture modeling in the following discussion. We will further show how the likelihood of a count-mixture model is related to that of an NRMI mixture by fixing the scale parameter of $G$ but treating $\beta$ as a random variable. % how We purposely  keep $\beta$ in the above  analysis to make connections to the NRMI mixtures. 

%cannot be absorbed into $G$.
   
   \subsection{Cluster Structures}\label{sec:thms}
   Below we show that  
   a count-mixture model is characterized by a compound Poisson process, under which the \emph{a priori} cluster structure  becomes evident. 
As in (\ref{eq:f_G5}), with $\beta\equiv 1$, we have
$
f(n_{1:K}|G) =\prod_{k=1}^K \mbox{Pois}(n_k;r_k).
$
With %$n_k:=\sum_{i=1}^m \delta(z_i=k)$ and
 $X:=\sum_{k=1}^{K} n_k\delta_{\omega_k}$, we obtain a Poisson process
$
 X\sim{\mbox{PP}}(G)
$
such that $X(A)\sim{\mbox{Pois}}(G(A))$ for each $A\subset\Omega$. With $G$ marginalized out,  $X$ becomes  a $G$ mixed Poisson process, %For each $A\subset\Omega$, 
with  characteristic functions %of $X$ % of $X(A)$ 
%can be expressed as
$
 \E[e^{iuX(A)}] = %\E[\E[e^{iuX(A)}|G(A)]] = 
 \E[ e^{G(A)(e^{iu}-1)}] = \exp\left\{ \int\hspace{-2mm}\int_{\mathbb{R}_+\times A}(e^{(e^{iu}-1)r}-1)\nu(dr d\omega)\right\} .\notag % =  \int\hspace{-2mm}\int_{\mathbb{R}_+\times A}(e^{(e^{iu}-1)r}-1)\nu(dr d\omega).\notag
$
% %Using (\ref{LevyCF}) and 
Since $e^{(e^{iu}-1)r}-1= % e^{-r}[(e^{re^{iu}}-1) - (e^r-1)] =
  \sum_{n=1}^\infty \frac{r^n e^{-r}}{n!}(e^{iun}-1)$, we have 
 \begin{align}\label{eq:CF0}
% &\E[e^{iuX(A)}] %&= \E[ e^{G(A)(e^{iu}-1)}] = \exp\left\{ \int\hspace{-2mm}\int_{\mathbb{R}_+\times A}(e^{(e^{iu}-1)r}-1)\nu(dr d\omega)\right\}\notag\\
  %\E[e^{iuX(A)}]  
 % \E[e^{iuX(A)}] = %\E[\E[e^{iuX(A)}|G(A)]] = 
% &\E[ e^{G(A)(e^{iu}-1)}] %= \exp\left\{ \int\hspace{-2mm}\int_{\mathbb{R}_+\times A}(e^{(e^{iu}-1)r}-1)\nu(dr d\omega)\right\} \notag\\ % =  \int\hspace{-2mm}\int_{\mathbb{R}_+\times A}(e^{(e^{iu}-1)r}-1)\nu(dr d
   \E[e^{iuX(A)}] & = \exp\left\{G_0(A)   \sum_{n=1}^\infty (e^{iun}-1) {\int_0^\infty \frac{r^n e^{-r}} {n!} \rho(dr)}\right\}\\ % \left[(e^{re^{iu}}-1) - (e^r-1)\right]\nu(dr d\omega)
   \label{eq:CF}
 &=\exp\left\{G_0(A)\int_{0}^\infty(1-e^{-r})\rho(dr)   \left(\sum_{n=1}^\infty e^{iun}\frac{\int_0^\infty r^n e^{-r} \rho(dr)}{n!\int_{0}^\infty(1-e^{-r})\rho(dr)} -1 \right)\right\}.% \left[(e^{re^{iu}}-1) - (e^r-1)\right]\nu(dr d\omega)
 \end{align}

\begin{thm}[Poisson Mixture Process]\label{thm:MixPois}
 %For the count-mixture model, with $G$ marginalized out, 
 $X:=\sum_{k=1}^{K} n_k\delta_{\omega_k}$ can be considered as a draw %from the Poisson process  
 %\beq
% X\sim\emph{\mbox{PP}}(G).
% \eeq
 %where $G$ is a homogenous completely random measure with L\'{e}vy measure $\nu(drd\omega) = \rho(dr) G_0(d\omega)$.
% Marginalizing out $G$ leads to 
from a $G$ mixed Poisson process, % $X\sim\E_G(\emph{\mbox{PP}}(G))$, 
with characteristic functions as in (\ref{eq:CF0}) and
%  \beq\label{eq:CF_PoisMix}
% \E[e^{iuX(A)}] = \E_G[ e^{G(A)(e^{iu}-1)}] = \exp\left\{\int\hspace{-2mm}\int_{\mathbb{R}_+\times A}(e^{(e^{iu}-1)r}-1)\nu(dr d\omega)\right\}
% \eeq
 L\'{e}vy measure
 \begin{align}\label{eq:LevyCountMixture}
 \tilde{\nu}(dnd\omega) &= \sum_{j=1}^\infty  \int_{0}^\infty \frac{ r^j e^{-r} } {j!}\rho(dr) \delta_j(dn) G_0(d\omega).\notag%\\
 %&=  \sum_{j=1}^\infty  \frac{\gamma_0^{-1}(-1)^n }{j!} \left.\frac{d^n\left(\ln\E[e^{-sG(\Omega)}]\right)}{ds^n}\right|_{s=1}   \delta_j(dn) G_0(d\omega).
 \end{align}
 %Since although a draw from $G$ may consists of countably infinite atoms, the total random mass $G(\Omega)$ is finite. Thus $X(\Omega)\sim\mbox{Pois}(G(\Omega))$ is finite and then the number of atoms if finite  
  \end{thm}

 \begin{thm}[Compound Poisson Process]\label{thm:compoundPoisson}
 It is evident from (\ref{eq:CF}) that the $G$ mixed Poisson process is also a compound Poisson process, a random draw of which can be expressed as 
  \begin{align}
 &X=\sum_{k=1}^l n_k \delta_{\omega_k},~l\sim\emph{\mbox{Pois}}\left(\gamma_0\int_{0}^\infty(1-e^{-r})\rho(dr)\right),~n_k\stackrel{iid}{\sim} \sum_{j=1}^\infty  \frac{ {\int_{0}^\infty r^j e^{-r} \rho(dr)} }{{j!}\int_{0}^\infty(1-e^{-r})\rho(dr)} \delta_j,\notag
 %(n_k,\omega_k) \sim \tilde{\nu}(dnd\omega)/\tilde{\nu}^+, 
 \end{align} 
 where $\int_{0}^\infty(1-e^{-r})\rho(dr)%\le \int_{0}^\infty\min\{1,r\}\rho(dr) 
 <\infty$ by definition and $\omega_k\stackrel{iid}{\sim} g_0$. % is the base distribution. 
  \end{thm}
  
 The compound Poisson representation dictates %in the prior 
the count-mixture model to have a Poisson distributed finite number of  clusters, whose sizes follow a positive discrete distribution %determined by $\rho(dr)$ as 
as $P(n=j|\rho)=\frac{ {\int_{0}^\infty r^j e^{-r} \rho(dr)} }{{j!}\int_{0}^\infty(1-e^{-r})\rho(dr)},~j=1,2,\cdots$. %, and the total number of samples $m:=X(\Omega)=\sum_{k=1}^l n_k$, which follows a Poisson mixture distribution by construction, also follows a compound Poisson distribution. 
 The mass parameter $\gamma_0$ has a linear relationship with the expected number of clusters, % and hence the expected number of samples, 
 but has no direct impact on the cluster-size distribution.  % , including both the cluster-number and cluster-size distributions, specified by the count-mixture model.
Since the cluster indexes are unordered and exchangeable, %We define $\mathcal{D}_m=\{\omega_k\}_{k:n_k>0}$ as the set of points of discontinuity observed among the $m$ data samples in the count-mixture model.
without loss of generality, in the following discussion,  we relabel %the indexes of the 
the atoms  in $\mathcal{D}_m$ %=\{\omega_k\}_{k:n_k>0}$ 
in order of appearance from $1$ to $l:=|\mathcal{D}_m|$ and then $z_i\in\{1,\cdots,l\}$ for $i=1,\cdots,m$,  $n_k>0$ if and only if $1\le k \le l$ and $n_k\equiv 0$ if $l<k\le K$. %Note that $l$ is a simplified notation of $|\mathcal{D}_m|$, which depends on %is a function of 
%the sample size $m$.

\begin{thm}[Exchangeable Cluster/Partition Probability Functions]\label{thm:ECPF}
The count-mixture model %,  with L\'{e}vy measure $\nu(drd\omega)=\rho(dr)G_0(d\omega)$, %shown in (\ref{eq:CRMcountandmixture})
has a fully factorized  exchangeable cluster probability function (ECPF) as
\begin{align}%\label{eq:ECPF_CRM}
f(\zv,m|\gamma_0,\rho) &= \E_{G} [f(\zv,m|G)] = \frac{\gamma_0^l}{m!} e^{\gamma_0\int_{0}^\infty(e^{-r}-1)\rho(dr)} %\E\left[e^{-G(\Omega\backslash\mathcal{D}_m) }\right]  
\prod_{k=1}^l \int_0^\infty r^{n_k} e^{-r} \rho(dr),\notag
\end{align}
a marginal distribution for the sample size $m=X(\Omega)$ with probability generating function (PGF)
$ % \begin{align}\label{eq:PGF_m}
 \E[t^{m}|\gamma_0,\rho] =   e^{ \gamma_0 \int_{0}^\infty (e^{-(1-t)r}-1)\rho(dr)} % = \E[e^{-(1-t)G(\Omega)}]
  %&=   e^{ -\gamma_0\int_{0}^\infty (1-e^{-r})\rho(dr)}  e^{ \gamma_0\int_{0}^\infty e^{-r}(e^{sr}-1)\rho(dr) }\\
  %&=   e^{ -\gamma_0\int_{0}^\infty (1-e^{-r})\rho(dr)}  \sum_{k=0}^\infty \frac{ \gamma_0^k [\int_{0}^\infty e^{-r}(e^{sr}-1)\rho(dr) ]^k}{k!} \\
  %&=   e^{ -\gamma_0\int_{0}^\infty (1-e^{-r})\rho(dr)}  \sum_{k=0}^\infty \frac{ \gamma_0^k [\int_{0}^\infty e^{-r}( \sum_{j=1}^\infty \frac{s^jr^j}{j!})\rho(dr) ]^k}{k!} 
$ % \end{align}
 and probability mass function (PMF) $f_M(m|\gamma_0,\rho)=\frac{d^m (\E[t^m|\gamma_0,\rho])} {d t^m } |_{t=0}$,
and an exchangeable partition probability function (EPPF) as
\begin{align}%\label{eq:EPPF_CRM}
f(\zv|m,\gamma_0,\rho) &= {f(\zv,m|\gamma_0,\rho) }\big/{f_M(m|\gamma_0,\rho)}.\notag %= \frac{{\gamma_0^l}\E\left[e^{-G(\Omega\backslash\mathcal{D}_m) }\right]  \prod_{k=1}^l  \int_0^\infty r_k^{n_k} e^{-r_k} \rho(dr)}{{m!}{f_M(m|\gamma_0,\rho)}},
% =&\frac{1}{N!}\E\left[e^{-G(\Omega\backslash\mathcal{D}) }\right] \notag\\ %\Big(\frac{\gamma_0}{\Gamma(1-a)}\Big)^{|\mathcal{D}|}
% &({\gamma_0})^{|\mathcal{D}|}\prod_{k:\omega_k\in\mathcal{D}} \Big\{%\int_{0}^\infty {r_k^{n_k-a-1} }e^{-\frac{r_k}{p}} d{r_k}
% \E[r_k^{n_k}e^{-r_k}]\int_{\Omega} \prod_{i:z_i=k} \kappa(x_i|\omega_{k}) \pi(d\omega_{k})\Big\},
\end{align}
%where  %$f_M(m):=\E_G[f(m|G)]$ is the marginal distribution of $m$ and 
%$\E\left[e^{-G(\Omega\backslash\mathcal{D}_m) }\right] = e^{-\gamma_0\int_{0}^\infty(1-e^{-r})\rho(dr)}$  and $\E\left[e^{-(1-s)G(\Omega\backslash\mathcal{D}_m) }\right] = e^{-\gamma_0\int_{0}^\infty(1-e^{(s-1)r})\rho(dr)}$. %  according to (\ref{LevyCF}).
\end{thm}

With $l^{-i}$ representing the number of clusters in $\zv^{-i}:=\zv\backslash z_i$ and $n_k^{-i}:=\sum_{j\neq i} \delta(z_j=k)$, we may reexpress the ECPF as
%\beq
%f(z_i,\zv^{-i},m|\gamma_0,\rho)= \frac{\gamma_0^{l^{-i}}}{m!} e^{\gamma_0\int_{0}^\infty(e^{-r}-1)\rho(dr)} 
%\Big(\gamma_0\int_0^\infty re^{-r}\rho(dr)\Big)^{\delta(z_i=l^{-i}+1)} \prod_{k=1}^{l^{-i}} \int_0^\infty r^{n_k^{-i}+\delta(z_i=k)} e^{-r} \rho(dr).\notag
%\eeq
\small
\beqs
&f(z_i,\zv^{-i},m|\gamma_0,\rho)=%\sum_{z_i}f(z_i,\zv^{-i},m|\gamma_0,\rho) = 
\frac{f(\zv^{-i},m-1|\gamma_0,\rho)}{m}\Big({\gamma_0\int_0^\infty re^{-r}\rho(dr)\delta(z_i=l^{-i}+1)+ \sum_{k=1}^{l^{-i}} \frac{\int_0^\infty r^{n_k^{-i}+1} e^{-r} \rho(dr)}{\int_0^\infty r^{n_k^{-i}} e^{-r} \rho(dr)}\delta(z_i=k)  }\Big).\notag
\eeqs\normalsize
Marginalized out  $z_i$ with $f(\zv^{-i},m|\gamma_0,\rho)=\sum_{z_i=1}^{l^{-i}+1}f(z_i,\zv^{-i},m|\gamma_0,\rho)$, we have
\small
\beqs
&f(\zv^{-i},m|\gamma_0,\rho)=%\sum_{z_i=1}^{l^{-i}+1}f(z_i,\zv^{-i},m|\gamma_0,\rho)= %\sum_{z_i}f(z_i,\zv^{-i},m|\gamma_0,\rho) = 
\frac{f(\zv^{-i},m-1|\gamma_0,\rho)}{m}\Big({\gamma_0\int_0^\infty re^{-r}\rho(dr)+ \sum_{k=1}^{l^{-i}} \frac{\int_0^\infty r^{n_k^{-i}+1} e^{-r} \rho(dr)}{\int_0^\infty r^{n_k^{-i}} e^{-r} \rho(dr)}  }\Big).\notag
\eeqs
\normalsize

\begin{thm}[Prediction Rule]\label{thm:predict} Since $f(z_i|\zv^{-i},m,\gamma_0,\rho) = \frac{f(z_i,\zv^{-i},m|\gamma_0,\rho)}{f(\zv^{-i},m|\gamma_0,\rho)}$, %where $f(\zv^{-i},m|\gamma_0,\rho)=\sum_{z_i=1}^{l^{-i}+1}f(z_i,\zv^{-i},m|\gamma_0,\rho)$, 
we can express the prediction rule of the count-mixture model %,  with L\'{e}vy measure $\nu(drd\omega)=\rho(dr)G_0(d\omega)$, %shown in (\ref{eq:CRMcountandmixture})
as
\beq%\label{eq:PredictRuleG}
P(z_{i} = k|\zv^{-i},m,\gamma_0,\rho) \propto
\begin{cases}
\frac{\int_{0}^\infty r^{n_k^{-i}+1} e^{-r} \rho(dr)}{\int_0^\infty r^{n_k^{-i}}e^{-r}  \rho(dr)} , & \emph{\mbox{for }} k=1,\cdots,l^{-i};\\
  \gamma_0\int_0^\infty re^{-r}\rho(dr), & \emph{\mbox{if } }k=l^{-i}+1.
\end{cases}\notag
\eeq
\end{thm}

We may use a Gibbs sampler  to simulate the EPPF $f(\zv|m,\gamma_0,\rho)$: %exchangeable random partitions $\Pi_m$ of $[m]$: % based on the EPPF $f(\zv|m,\gamma_0,\rho)$: 
initializing $\zv=(z_1,\cdots,z_m)$ at random, in each Gibbs sampling iteration we remove  elements one by one from their current clusters and immediately reassign the removed element to a cluster based on the prediction rule in Theorem \ref{thm:predict}; as the Markov chain converges, the ties between the elements of $\zv$ 
in each iteration would define an exchangeable random partition $\Pi_m$ of $[m]$, and the ties between a subset of $j$ elements uniformly at random subsampled without replacement from $\zv$ would define a size-dependent  exchangeable random partition $(\Pi_j|m)$.  %We may construct an exchangeable random partition $\zv$that 

\begin{cor}[Size-Dependent EPPF]\label{sdEPPF}
Adding an element to a sample of size $m$, the EPPF of $\zv=(z_1,\cdots,z_m)$ in a sample of size $m+1$ is revised from the original EPPF $f(\zv|m,\gamma_0,\rho)$  as %a function of the EPPF as
\beqs
&f(\zv|m+1,\gamma_0,\rho) = f(\zv|m,\gamma_0,\rho)\frac{f_{M}(m|\gamma_0,\rho)}{(m+1) f_{M+1}(m+1|\gamma_0,\rho)}\Big(\gamma_0\int_0^\infty re^{-r}\rho(dr)+ \sum_{k=1}^{l}\frac{\int_{0}^\infty r^{n_k+1} e^{-r} \rho(dr)}{\int_{0}^\infty r^{n_k} e^{-r} \rho(dr)}\Big)\notag
\eeqs
and %$z_{1:m+1}=(\zv,z_{m+1})$ can be constructed %by  assigning element $m$ is assigned to $A_{z_{m+1}}$ 
the new element is clustered
with the prediction rule $f(z_{m+1}|\zv,m+1,\gamma_0,\rho)=\frac{f(\zv,z_{m+1},m+1|\gamma_0,\rho)}{f(\zv,m+1|\gamma_0,\rho)}$.
\end{cor}

\begin{cor}[Equivalent Construction] \label{cor:EPPF_to_ECPF}
If both the marginal distribution $f_M(m|\gamma_0,\rho)$ and  the EPPF $f(\zv|m,\gamma_0,\rho)$ are  known, then the count-mixture model can also be constructed as 
\beqs%\label{eq:CRMcountandmixture1}
x_i\sim\kappa(\omega_{z_i}),~\omega_k\sim g_0,~ \zv\sim f(\zv|m,\gamma_0,\rho),~m\sim f_M(m|\gamma_0,\rho). \notag %f(z_i|G)={r_{z_i}}/{G(\Omega)},
\eeqs
\end{cor}

There are several notable distinctions differing a count-mixture model from an NRMI mixture model. First, the model introduces a new mechanism to generate a sample of arbitrary size %by treating the sample size $m$ as a random variable, the model 
and specifies a prior distribution on all possible exchangeable random partitions. Second, both the conditional likelihood and the ECPF are fully factorized by construction and hence naturally amenable to posterior simulation. Third, $G$ no longer scales freely, removing the redundancy between its parameters, and $K=\infty$ is no longer mandatory. % $G(\Omega)$ can be zero with a nonzero probability. % of $G$. % are no longer redundant. 
Fourth, %$P(\Pi_m|m)=P(\Pi_m|m+1)$
$f(\zv|m,\gamma_0,\rho)=f(\zv|m+1,\gamma_0,\rho)$ is not required % $f(\zv^{-i}|m,\gamma_0,\rho) = f(\zv^{-i}|m-1,\gamma_0,\rho)$ 
and the EPPF may not satisfy the addition rule in~(\ref{eq:addrule}). % and hence it might be an inconsistent EPPF with inconsistent %, whose 
%exchangeable random partitions. % are not consistent in distribution as $m$ varies. 
Fifth, the ECPF, EPPF and prediction rule are all straightforward to calculate, without the need of introducing an auxiliary  variable. 
Last but not least, the count-mixture model is characterized by a %completely random measure mixed Poisson process, which is also a 
compound Poisson process, %the analysis of which  often leads to unique insights about the model properties, including  
which clearly specifies 
the \emph{a priori} cluster-number and cluster-size  distributions. % with $m$ known or unknown. 

\section{Generalized Negative Binomial Processes}\label{sec:gNBP} % and Chinese restaurant processes}

The previous section shows that every 
%A cluster structure can be generated from a
completely random measure $G$ with a finite total random mass  can  generate a  cluster structure, 
with the $a~priori$ cluster-number and cluster-size distributions determined by the L\'{e}vy measure. In the following discussion, 
we study the generalized NB process (gNBP) count-mixture model where $G\sim\mbox{gGaP}(G_0,a,(1-p)/p)$  with $a<0$, $a=0$ or $0<a<1$, for which both $\int_{0}^\infty r^je^{-r}\rho (dr)$ and  $\int_{0}^\infty (1-e^{-r})\rho (dr)$ are convenient to calculate. %have simple analytic expressions.

\subsection{Generalized and Truncated Negative Binomial Distributions}

Marginalizing out $\lambda$ in %a generalized gamma-Poisson mixture distribution
$m\sim\mbox{Pois}(\lambda),~\lambda\sim{{}}\mbox{gGamma}(\gamma_0,a,p/(1-p))$ leads to a generalized NB distribution $m\sim{{}}\mbox{gNB}(\gamma_0,a,p)$, with shape parameter $\gamma_0$, discount parameter $a<1$ and probability parameter $p$, whose PGF is
$
\E[t^m]  = \E[\E[t^m|\lambda]] = e^{-\frac{\gamma_0((1-pt)^a-(1-p)^a)}{ap^a}}, %=e^{\gamma_0\frac{(1-p)^a}{ap^a}} \sum_{k=0}^\infty \frac{1}{k!} {\left(\frac{-\gamma_0}{ap^a}\right)^k} \sum_{j=0}^\infty \binom{ak}{j}(-pz)^j.
$
mean is $\gamma_0\big(\frac{p}{1-p}\big)^{1-a}$ and variance is $\gamma_0\big(\frac{p}{1-p}\big)^{1-a}\frac{1-ap}{1-p}$. The PGF of this distribution was presented in \citet{willmot1988remark,gerber1992generalized}. With the PGF reexpressed as $\E[t^m] =e^{\gamma_0\frac{(1-p)^a}{ap^a}}$ $\sum_{k=0}^\infty \frac{1}{k!} {\left(\frac{-\gamma_0(1-pt)^a}{ap^a}\right)^k} =e^{\gamma_0\frac{(1-p)^a}{ap^a}}$ $\sum_{k=0}^\infty \frac{1}{k!} {\left(\frac{-\gamma_0}{ap^a}\right)^k} \sum_{j=0}^\infty \binom{ak}{j}(-pt)^j$,
we derive the PMF as 
\beqs\label{eq:f_M}
f_M(m|\gamma_0,a,p)
 = %e^{{\gamma}\frac{(1-p)^a}{ap^a}}(-p)^m \sum_{k=0}^\infty \frac{1}{k!}{\left(\frac{-\gamma}{ap^a}\right)^k}  \binom{ak}{m} =
\frac{p^m}{m!}e^{{\gamma_0}\frac{(1-p)^a}{ap^a}} \sum_{k=0}^\infty \frac{1}{k!}{\left(-\frac{\gamma_0}{ap^a}\right)^k} \frac{\Gamma(m-ak)}{\Gamma(-ak)}, ~m=0,1,\cdots.
%\\
%&= \frac{\Gamma(m-a+\gamma_0p^{-a})}{m!\Gamma(1-a+\gamma_0p^{-a})}\gamma_0e^{-\gamma_0\frac{(1-(1-p)^a)}{ap^{a}}}p^{m-a}
%&=\exp\left({\gamma}\frac{(1-p)^a}{ap^a}\right)(-p)^m  \sum_{k=0}^\infty \frac{\left(\frac{-\gamma}{ap^a}\right)^k}{k!}  (ak)(ak-1)\cdots(ak-m+1)
\eeqs %\end{align}

Since the PGF  can be reexpressed as $\E[t^m]=e^{\frac{\gamma0(1-(1-p)^a)}{ap^a}\left(\frac{1-(1-pt)^a}{1-(1-p)^a}-1\right)}$, the generalized NB distribution
$m\sim{{}}\mbox{gNB}(\gamma_0,a,p)$ can also be generated from a compound Poisson distribution as
$
m=\sum_{i=1}^l u_i, u_i\stackrel{iid}{\sim}\mbox{TNB}(a,p),~l\sim\mbox{Pois}\big(\frac{\gamma_0(1-(1-p)^a)}{ap^a}\big),
$
where $u\sim\mbox{TNB}(a,p)$ is a truncated NB distribution, with PGF $\E[t^{u}] =  %\sum_{n=1}^\infty  \frac{\Gamma(n-a)}{n!\Gamma(-a)}\frac{p^n(1-p)^{-a}}{1-(1-p)^{-a}} z^n =
\frac{1-(1-pt)^a}{1-(1-p)^a}$ and thus $\E[u] = \frac{a(1-p)^a}{1-(1-p)^a}\frac{p}{1-p}$ %, $\mbox{Var}[u]=(\frac{1-ap}{1-p}-\E[u])\E[u]$ 
and PMF
\beqs\label{eq:TNB}
f_U(u|a,p)= \frac{\Gamma(u-a)}{u!\Gamma(-a)}\frac{p^u(1-p)^{-a}}{1-(1-p)^{-a}},~u=1,2,\cdots.
\eeqs
Note that as $a\rightarrow 0$, $u\sim\mbox{TNB}(a,p)$ becomes a logarithmic distribution \citep{LogPoisNB} $u\sim\mbox{Log}(p)$ with PMF $f_U(u|p)=\frac{-1}{\ln(1-p)}\frac{p^u}{u}$, $u=1,2,\cdots$, %\citep{johnson2005univariate} 
and $m\sim\mbox{gNB}(\gamma_0,a,p)$ becomes a NB distribution $m\sim\mbox{NB}(\gamma_0,p)$. The truncated NB distribution with $0<a<1$ is the extended NB distribution introduced in \citet{engen1974species}.

Denote $\sum_{\sum_{k=1}^l n_k = m}$ as the summation over all sets of positive integers $(n_1,\cdots,n_l)$ with ${\sum_{k=1}^l n_k = m}$. Using both (\ref{eq:TNB}) %and PGF of the truncated NB distribution 
and $
\left[\frac{1-(1-pt)^a}{1-(1-p)^a}\right]^{l}= \frac{ \sum_{k=0}^l \binom{l}{k} (-1)^k \sum_{j=0}^\infty\binom{ak}{j} (-pt)^j }{  [1-(1-p)^a]^l}
$,  we may express the PMF of the sum-truncated NB distribution $m=\sum_{i=1}^l u_i, u_i\stackrel{iid}{\sim}\mbox{TNB}(a,p)$ %in two different ways 
as
$f_M(m|l,a,p) = 
\sum_{\sum_{k=1}^l n_k=m} \prod_{k=1}^l {\frac{\Gamma(n_k-a)}{n_k!\Gamma(-a)} \frac{p^{n_k}(1-p)^{-a}}{1-(1-p)^{-a}}}%\notag\\
=\frac{p^m}{ [1-(1-p)^a]^l} {\sum_{k=0}^l (-1)^k \binom{l}{k}  \frac{\Gamma(m-ak)}{m!\Gamma(-ak)}  },%= \frac{p^{m}}{ [ \frac{1-(1-p)^{a}}{a}]^{l}} \sum_{\sum_{k=1}^l n_k=m} \prod_{k=1}^l \frac{\Gamma(n_k-a)}{n_k!\Gamma(1-a)}
 $ leading to identity 
 \begin{align}\label{eq:identity1}
%\frac{p^m \sum_{k=0}^l (-1)^k \binom{l}{k}  \frac{\Gamma(m-ak)}{m!\Gamma(-ak)}  }{ [1-(1-p)^a]^l}\frac{ [\Gamma(-a) ((1-p)^{a}-1)]^{l}} {p^{m}}  = \sum_{\sum_{k=1}^l n_k=m} \prod_{k=1}^l \frac{\Gamma(n_k-a)}{n_k!}\\
S_a(m,l) := \frac{m!}{l!}\sum_{\sum_{k=1}^l n_k=m} \prod_{k=1}^l \frac{\Gamma(n_k-a)}{n_k!\Gamma(1-a)} = \frac{1}{l!a^{l}}\sum_{k=0}^l (-1)^k \binom{l}{k}  \frac{\Gamma(m-ak)}{\Gamma(-ak)},
\end{align}
%whose right-hand side is easier to calculate. 
where
%the right-hand side of which can be written as 
$S_a(m,l)$ are %Toscano's formula or 
generalized Stirling numbers of the first kind, which can be recursively calculated with $S_a(m,1)=\frac{\Gamma(m-a)}{\Gamma(1-a)}$, $S_a(m,m)=1$ and $S_a(m+1,l) = (m-al)S_a(m,l)+S_a(m,l-1)$  \citep{charalambides2005combinatorial,csp}.
%Note that 
When $a\rightarrow0$, %$u_i\sim{\mbox{TNB}}(a,p)$ becomes a logarithmic distribution $u_i\sim{\mbox{Log}}(p)$,  
the sum-truncated NB distribution becomes the sum-logarithmic distribution \citep{NBP2012}, and (\ref{eq:identity1}) becomes %(\ref{eq:identity}).  
$ %\beq\label{eq:identity}
\frac{m!}{l!}\sum_{\sum_{k=1}^l n_k = m}\prod_{k=1}^l \frac{1}{n_k} =|s(m,l)|,
$ where $|s(m,l)|=S_0(m,l)$ are unsigned Stirling numbers of the first kind \citep{johnson2005univariate}. %,charalambides2005combinatorial}.%\eeq

\subsection{Model Properties} %Generalized Negative Binomial and Chinese Restaurant Processes}
\label{sec:GNBP}

Marginalizing out $G$ in the generalized gamma process mixed Poisson process 
\beq\label{eq:gGaPP0}
X\sim\mbox{PP}(G),~G\sim{{}}\mbox{gGaP}(G_0,a,{(1-p)}/{p})
\eeq %\eeqs %in (\ref{eq:gGaPP}) 
leads to a generalized NB process
$
X\sim\mbox{gNBP}(G_0,a,p),
$
such that for each $A\subset \Omega$, $X(A)\sim\mbox{gNB}(G_0(A),a,p)$. 
Since $\rho(dr) = \frac{r^{-a-1}}{\Gamma(1-a)}e^{-\frac{1-p}{p}r}dr$, we have $\int_{0}^\infty r^{n}e^{-r}\rho(dr)= {\frac{\Gamma(n-a)}{{\Gamma(1-a)} }p^{n-a}}$ and $\int_0^\infty(1-e^{-r})\rho(dr) = \frac{1-(1-p)^a}{ap^{a}}$.
Based on the analysis in Section \ref{sec:thms}, the L\'{e}vy measure of $X$ %\sim\mbox{gNBP}(G_0,a,p)$ 
can be expressed as
$
\tilde{\nu}(dnd\omega)=\sum_{m=1}^\infty \frac{\Gamma(m-a)}{m!\Gamma(1-a)}{p^{m-a}} \delta_m(dn) {G_0(d\omega)},\notag
%\sum_{m=0}^\infty\frac{\Gamma(m-a)}{m!\Gamma(-a)}p^m(1-p)^{-a}\delta_m(dn) {G_0(d\omega)}.
$
and the gNBP count-mixture model shown in (\ref{eq:CRMcountandmixture}), with $\beta=1$ and $G\sim\mbox{gGaP}(G_0,a,(1-p)/p)$, is characterized by a compound Poisson process as
$ %\begin{align}%\label{eq:GNBPdraw}
X=\sum_{k=1}^{l} n_k\delta_{\omega_k},~l\sim\mbox{Pois}\Big(\gamma_0\frac{(1-(1-p)^a)}{ap^a}\Big),~n_k  \stackrel{iid}{\sim} \mbox{TNB}(a,p),~\omega_k \stackrel{iid}{\sim} g_0 \notag%(n_k, \omega_k)\stackrel{iid}{\sim} \mbox{TNB}(n_k;a,p) g_0(\omega_k). %,~k=1,\cdots,K,
$.  

The ECPF of the gNBP %generalized NB process 
count-mixture
 model % shown in (\ref{eq:CRMcountandmixture}) %, where $\beta=1$ and $G\sim\mbox{gGaP}(G_0,a,(1-p)/p)$, 
 can be expressed as
\beqs \label{eq:f_Z_M}
f(\zv,m|\gamma_0,a,p) %&= \E_{G}[ f(\zv,m|G) ]%= %\frac{1}{N!}\E\left[e^{-G(\Omega\backslash\mathcal{D}) }\right]  \prod_{k:\omega_k\in\mathcal{D}} \gamma_0 \int r_k^{n_k} e^{-r_k} \rho(dr)\notag\\
%=\frac{1}{m!}\E\left[e^{-G(\Omega\backslash\mathcal{D}) }\right]  \prod_{k:\omega_k\in\mathcal{D}} \gamma_0 \int r_k^{n_k} e^{-r_k} \rho(dr)\notag\\
=\frac{1}{m!}e^{-\frac{\gamma_0(1-(1-p)^a)}{ap^{a}}}
 \gamma_0^{l{}} p^{m-al{}}
\prod_{k=1}^{l{}} \frac{\Gamma(n_k-a)}{{\Gamma(1-a)} }.
\eeqs 
Under the fully factorized ECPF,  it is clear that $\gamma_0\sim\mbox{Gamma}(e_0,1/f_0)$ is a conjugate prior % the conjugate prior for $\gamma_0$ is a gamma distribution, 
and the other two univariate parameters $a$ and $p$ can also be conveniently inferred. %, e.g., using % using algorithms such as the 
%the griddy-Gibbs  sampler \citep{griddygibbs}.
The EPPF %of the generalized NB process 
is the ECPF in (\ref{eq:f_Z_M}) divided by the marginal distribution of $m$ in (\ref{eq:f_M}), expressed as
\begin{align} \label{eq:EPPF}
 f(\zv|m,\gamma_0,a,p) &= %\frac{ f(\zv,m|\gamma_0,a,p)}{\mbox{gNB}(m;\gamma_0,a,p)} = 
 p_m(n_1,\cdots,n_{l{}}) % \notag\\ %\frac{f(\zv,m|\gamma_0,a,p) }{f(m|G_0,a,p) } =
=\frac{1}{e^{\frac{\gamma_0}{ap^{a}}}  \sum_{k=0}^\infty \frac{1}{k!}{\left(-\frac{\gamma_0}{ap^a}\right)^k}\frac{\Gamma(m-ak)}{\Gamma(-ak)} }  \gamma_0^{l{}}  p^{-al{}} %\prod_{k:\omega_k\in\mathcal{D}}
\prod_{k=1}^{l{}}\frac{\Gamma(n_k-a)}{\Gamma(1-a)}.
%\\
%&= \frac{\Gamma(1-a+\gamma_0p^{-a}) \gamma_0^{l-1}  p^{-a(l-1)}}{ \Gamma(m-a+\gamma_0p^{-a}) }  \prod_{k=1}^l\frac{\Gamma(n_k-a)}{\Gamma(1-a)}.
%\frac{1}{m! \sum_{k=0}^\infty \frac{\left(\frac{-\gamma}{ap^a}\right)^k}{k!} (-1)^m(-ak)_m} { \gamma_0^l p^{-al}}e^{-\frac{\gamma_0}{ap^{a}}}\prod_{k=1}^l\frac{\Gamma(n_k-a)}{\Gamma(1-a)}
\end{align}
If $a\rightarrow 0$, we  recover from (\ref{eq:EPPF}) the Ewens sampling formula $
f(\zv|\gamma_0,m)=%\E_{\widetilde{G}}[f(\zv|\widetilde{G},m)]=p(n_1,\cdots,n_{l{}})=
\frac{\gamma_0^{l{}}\Gamma(\gamma_0)}{\Gamma(m+\gamma_0)}\prod_{k=1}^{l{}}\Gamma(n_k)$, which is the EPPF of the CRP \citep{aldous:crp}.
We define the stochastic process with the EPPF in (\ref{eq:EPPF})  %in (\ref{eq:EPPF}) % probability function
as the generalized CRP, expressed as $X\sim{\mbox{gCRP}}(m,\gamma_0,a,p)$,
whose prediction rule is % can be expressed as
\beq\label{eq:PredictRule}
%p(z_{m+1} = k|\zv,m,\gamma_0,a,p)
P(z_{i} = k|\zv^{-i},m,\gamma_0,a,p)  \propto
\begin{cases}
n_k^{-i} -a, &  {\mbox{for }} k=1,\cdots,l^{-i};\\ %\frac{n_k-a}{m+\gamma_0p^{-a}-al}, 
\gamma_0 p^{-a}, & {\mbox{if } }k=l^{-i}+1; %\frac{\gamma_0 p^{-a}} {m+\gamma_0p^{-a}-al}, 
\end{cases}
\eeq
based on which we can construct a Gibbs sampler to draw exchangeable random partitions $\Pi_m$ of $[m]$ and hence size-dependent exchangeable random partitions  $(\Pi_j|m)$, for $1\le j\le m-1$.
Using $[z_i,z_j]$ to denote ties that $z_i=z_j$, with (\ref{eq:EPPF}), we have
$
 %p_2(1,1)
 %f(z_1\neq z_2|2)
 f(z_1,z_2|2)
 = \frac{\gamma_0p^{-a}}{(1-a)+\gamma_0p^{-a}},
 $
 $
 % f(z_1\neq (z_2= z_3)|3)=f((z_1=z_2)\neq  z_3|3) = f((z_1=z_3)\neq  z_2|3)
  f(z_1,[z_2,z_3]|3)
  %f(z_1,z_{2,3}|3)
  %f([z_1,z_2], z_3|3) 
  = f([z_1,z_3], z_2|3)
 %p_3(1,2)=p_3(2,1)
 = \frac{(1-a)\gamma_0p^{-a}}{(2-a)(1-a)+3(1-a)\gamma_0p^{-a}+\gamma_0^2p^{-2a}}
 $ and  $
 %p_3(1,1,1)
% f(z_1\neq z_2\neq z_3|3)
  f(z_1,z_2,z_3|3)= 
   \frac{\gamma_0^2p^{-2a}}{(2-a)(1-a)+3(1-a)\gamma_0p^{-a}+\gamma_0^2p^{-2a}}
 $. %One may readily show that
%It immediately follows that, 
If $a\neq0$, then the % the generalized CRP has  an inconsistent 
EPPF violates the addition rule in (\ref{eq:addrule}) since %is inconsistent since
$f(z_1,z_2|2) \neq f(z_1,z_2|3)$, where $f(z_1,z_2|3)= f(z_1,z_2, z_3|3) +f(z_1,[z_2,z_3]|3) +f([z_1,z_3], z_2|3)$.

  \begin{figure}[!tb]
\begin{center}
\includegraphics[width=0.96\columnwidth]{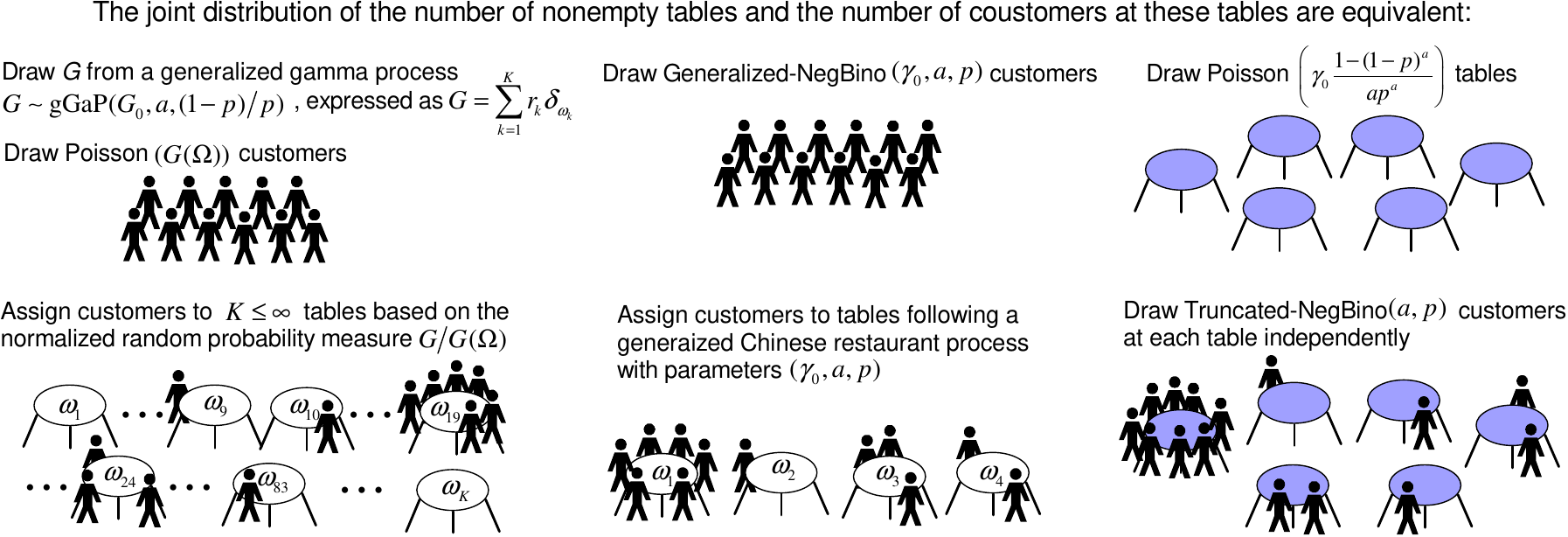}
\end{center}
\vspace{-6mm}
\caption{ \label{fig:gNBPdraw} \small The generalized NB process count-mixture model can be either constructed by assigning $\mbox{Pois}(G(\Omega))$ number of customers to tables following a normalized generalized gamma process $G/G(\Omega)$, where $G\sim\mbox{gGaP}(G_0,a,(1-p)/p)$, or constructed by assigning $\mbox{gNB}(\gamma_0,a,p)$ number of customers to tables following a generalized Chinese restaurant process $X\sim\mbox{gCRP}(\gamma_0,a,p)$, where $\gamma_0=G_0(\Omega)$. The joint distribution of  the number of clusters and the sizes of unordered clusters can be equivalently generated by first drawing $\mbox{Pois}\big(\gamma_0\frac{(1-(1-p)^a)}{ap^a}\big)$ number of tables, and then drawing  $\mbox{TNB}(a,p)$ number of customers independently at each table. 
}%\vspace{-2mm}
\end{figure}

With Corollary \ref{cor:EPPF_to_ECPF}, the gNBP count-mixture model in (\ref{eq:CRMcountandmixture}) %based on the generalized negative binomial process  $X\sim{{}}{\mbox{gNBP}}(G_0,a,p)$ can also be constructed as
%shown in (\ref{eq:gNBPcountmixture}) 
can also be constructed as
\beqs
x_i\sim\kappa(\omega_{z_i}),~\omega_k\sim g_0,~ \zv \sim\mbox{gCRP}(m,\gamma_0,a,p), ~m\sim{\mbox{gNB}}(\gamma_0,a,p).
 \eeqs
 As shown in Figure \ref{fig:gNBPdraw}, they both lead to the same cluster structure where
 the probability for a collection of unordered cluster sizes  $\{n_k\}_{1,l}$ %, generated by a random draw from the generalized NB process, 
 can be expressed as
\begin{align}\label{eq:ClusterStructure}
f(\{n_k\}_{1,l},m | \gamma_0,a,p)&=\delta_{\sum_{k=1}^l n_k}(m)\frac{1}{l!}\frac{m!}{\prod_{k=1}^l{n_k}}f(\zv,m|\gamma_0,a,p) \notag\\%&= \E_{G}[ f(\zv,m|G) ]%= %\frac{1}{N!}\E\left[e^{-G(\Omega\backslash\mathcal{D}) }\right]  \prod_{k:\omega_k\in\mathcal{D}} \gamma_0 \int r_k^{n_k} e^{-r_k} \rho(dr)\notag\\
%=\frac{1}{m!}\E\left[e^{-G(\Omega\backslash\mathcal{D}) }\right]  \prod_{k:\omega_k\in\mathcal{D}} \gamma_0 \int r_k^{n_k} e^{-r_k} \rho(dr)\notag\\
&=\delta_{\sum_{k=1}^l n_k}(m) \mbox{Pois}\Big(l;\gamma_0\frac{(1-(1-p)^a)}{ap^a}\Big) \prod_{k=1}^l\mbox{TNB}(n_k;a,p).
%&=\frac{e^{-\frac{\gamma_0(1-(1-p)^a)}{ap^{a}}} 
% \gamma_0^{l{}} p^{m-al{}}}{l!}
%\prod_{k=1}^{l{}} \frac{\Gamma(n_k-a)}{n_k!{\Gamma(1-a)} }.
\end{align}
Using %Corollary \ref{cor:table}, 
the EPPF  in (\ref{eq:EPPF}) and the identity in (\ref{eq:identity1}), the conditional distribution of the number of clusters $l$ in a sample of size $m$ %conditioning on the sample size $m$ 
can be expressed as
\begin{align}\label{eq:f_L2}
f_L(l|m,\gamma_0,a,p)& = \frac{1}{l!}\sum_{\sum_{k=1}^l n_k=m}\frac{m!}{\prod_{k=1}^l n_k}f(\zv|m,\gamma_0,a,p)%\notag\\
  =\frac{ \gamma_0^l p^{-al} S_a(m,l)}{e^{\frac{\gamma_0}{ap^{a}}}   \sum_{k=0}^\infty \frac{1}{k!} {\left(\frac{-\gamma_0}{ap^a}\right)^k}  \frac{\Gamma(m-ak)}{\Gamma(-ak)}} ,
\end{align}
which, since $\sum_{l=0}^mf_L(l|m,\gamma_0,a,p)=1$, further leads to identity 
\beq
e^{\frac{\gamma_0}{ap^{a}}}   \sum_{k=0}^\infty \frac{1}{k!} {\left(\frac{-\gamma_0}{ap^a}\right)^k}  \frac{\Gamma(m-ak)}{\Gamma(-ak)} = \sum_{l=0}^m \gamma_0^l p^{-al} S_a(m,l). \notag
\eeq
%This identity can be used to 
Thus we may simplify %the expressions of
  %$f_M(m|\gamma_0,a,p)$ in 
  (\ref{eq:f_M}) as 
$
f_M(m|\gamma_0,a,p)
 = %e^{{\gamma}\frac{(1-p)^a}{ap^a}}(-p)^m \sum_{k=0}^\infty \frac{1}{k!}{\left(\frac{-\gamma}{ap^a}\right)^k}  \binom{ak}{m} =
\frac{p^m}{m!}e^{-{\gamma_0}\frac{1-(1-p)^a}{ap^a}} \sum_{l=0}^m \gamma_0^l p^{-al} S_a(m,l)
%\sum_{k=0}^\infty \frac{1}{k!}{\left(-\frac{\gamma_0}{ap^a}\right)^k} \frac{\Gamma(m-ak)}{\Gamma(-ak)}, ~m=0,1,\cdots.
$, %the EPPF in %$f(\zv|m,\gamma_0,a,p)$ 
(\ref{eq:EPPF}) as
$
 f(\zv|m,\gamma_0,a,p) = %\frac{ f(\zv,m|\gamma_0,a,p)}{\mbox{gNB}(m;\gamma_0,a,p)} = 
% p_m(n_1,\cdots,n_{l{}}) % \notag\\ %\frac{f(\zv,m|\gamma_0,a,p) }{f(m|G_0,a,p) } =
\frac{\gamma_0^{l}  p^{-al}}{
\sum_{l=0}^m \gamma_0^l p^{-al} S_a(m,l)}
%e^{\frac{h_0}{a(1-p)^{a}}}  \sum_{k=0}^\infty \frac{1}{k!}{\big(-\frac{h_0}{a(1-p)^a}\big)^k}\frac{\Gamma(m-ak)}{\Gamma(-ak)} }  %\prod_{k:\omega_k\in\mathcal{D}}
\prod_{k=1}^{l{}}\frac{\Gamma(n_k-a)}{\Gamma(1-a)}
%\\
%&= \frac{\Gamma(1-a+\gamma_0p^{-a}) \gamma_0^{l-1}  p^{-a(l-1)}}{ \Gamma(m-a+\gamma_0p^{-a}) }  \prod_{k=1}^l\frac{\Gamma(n_k-a)}{\Gamma(1-a)}.
%\frac{1}{m! \sum_{k=0}^\infty \frac{\left(\frac{-\gamma}{ap^a}\right)^k}{k!} (-1)^m(-ak)_m} { \gamma_0^l p^{-al}}e^{-\frac{\gamma_0}{ap^{a}}}\prod_{k=1}^l\frac{\Gamma(n_k-a)}{\Gamma(1-a)}
$ and (\ref{eq:f_L2}) as $
 f(l|m,\gamma_0,a,p) = \frac{\gamma_0^{l}  p^{-al}S_a(m,l)}{
\sum_{l=0}^m \gamma_0^l p^{-al} S_a(m,l)}$.

 It is evident from (\ref{eq:ClusterStructure}) that
 the gNBP can adjust both $a$ and $p$ to fit the cluster sizes with $n_k\stackrel{iid}{\sim}  \mbox{TNB}(a,p)$,
 %$\{n_k\}_{1,l}$  %$\omega_k\in\mathcal{D}$
  %with a truncated NB distribution, 
  and adjust $\gamma_0$, $a$ and $p$ to fit the number of clusters $l$ with a Poisson distribution and the sample size with $m\sim \mbox{gNB}(\gamma_0,a,p)$. 
 The way the number and sizes of clusters %number of clusters and the sizes of these clusters 
 are modeled in the gNBP is unique, clearly explaining the roles of the three free model parameters on controlling %count distributions and hence 
 the cluster structure.

\vspace{-2mm}
\subsubsection{Reparameterized Generalized Negative Binomial Process}
\vspace{-2mm}
Since if a generalized gamma process 
$G\sim\mbox{gGap}(G_0,a,(1-p)/p)$ is normalized to construct a random probability measure, its scale and mass parameters are redundant to each other and hence it is standard to fix the scale parameter as $p/(1-p)=1$.  
In the gNBP, % count-mixture modeling, 
$\beta$ is fixed as one and $p$ serves as the probability parameter of the cluster-size distribution. % and thus shall not be simply fixed. 
To make connections to the approaches where $p$ could be  fixed, including the normalized generalized gamma process mixture model,  %we no longer fixes $\beta=1$ but fixes the scale parameter of the completely random measure as one: 
we modify the count-mixture model in (\ref{eq:CRMcountandmixture}) with $m\sim\mbox{Pois}(\beta H(\Omega)),~H\sim\mbox{gGaP}( H_0,a,1)$, where $\beta^a H_0 = G_0$, $h_0=H_0(\Omega) = \gamma_0\beta^{-a}$, $\beta: = \frac{p}{1-p}$ and $0<p<1$. The generalized gamma process mixed Poisson process in (\ref{eq:gGaPP0}) is now reparameterized as
\beq\label{eq:gGaPP}
X\sim\mbox{PP}(\beta H),~H\sim\mbox{gGaP}( H_0,a,1),
\eeq
in which the marginalization of $\beta H\sim \mbox{gGaP}\big( \big(\frac{p}{1-p}\big)^a H_0,a,\frac{1-p}{p}\big)$  leads to %a reparameterized gNBP
$
X\sim\mbox{gNBP}\big(\big(\frac{p}{1-p}\big)^a H_0,a,p\big).
$
The reparameterized gNBP count-mixture model has reparameterized ECPF and EPPF
\begin{align}  \label{eq:f_Z_M2}
f(\zv,m|h_0,a,p) %&= \E_{G}[ f(\zv,m|G) ]%= %\frac{1}{N!}\E\left[e^{-G(\Omega\backslash\mathcal{D}) }\right]  \prod_{k:\omega_k\in\mathcal{D}} \gamma_0 \int r_k^{n_k} e^{-r_k} \rho(dr)\notag\\
%=\frac{1}{m!}\E\left[e^{-G(\Omega\backslash\mathcal{D}) }\right]  \prod_{k:\omega_k\in\mathcal{D}} \gamma_0 \int r_k^{n_k} e^{-r_k} \rho(dr)\notag\\
&=\frac{p^m}{m!}e^{-\frac{h_0(1-(1-p)^a)}{a(1-p)^a}}
 h_0^{l}  (1-p)^{-al}
\prod_{k=1}^{l{}} \frac{\Gamma(n_k-a)}{{\Gamma(1-a)} },\\
\label{eq:EPPF_H}
 f(\zv|m,h_0,a,p) &= %\frac{ f(\zv,m|\gamma_0,a,p)}{\mbox{gNB}(m;\gamma_0,a,p)} = 
% p_m(n_1,\cdots,n_{l{}}) % \notag\\ %\frac{f(\zv,m|\gamma_0,a,p) }{f(m|G_0,a,p) } =
\frac{h_0^{l}  (1-p)^{-al}}{
\sum_{l=0}^m h_0^l (1-p)^{-al} S_a(m,l)}
%e^{\frac{h_0}{a(1-p)^{a}}}  \sum_{k=0}^\infty \frac{1}{k!}{\big(-\frac{h_0}{a(1-p)^a}\big)^k}\frac{\Gamma(m-ak)}{\Gamma(-ak)} }  %\prod_{k:\omega_k\in\mathcal{D}}
\prod_{k=1}^{l}\frac{\Gamma(n_k-a)}{\Gamma(1-a)},
\end{align}
which are the same as the ECPF in (\ref{eq:f_Z_M})  and EPPF in (\ref{eq:EPPF}), respectively, except that $\gamma_0$ is replaced with $h_0 \big(\frac{p}{1-p}\big)^a$.
As the reparameterized  EPPF matches that of  $X\sim{\mbox{gCRP}}(m,h_0,a,1-p)$,     the reparameterized gNBP count-mixture model 
can also be constructed as
\beqs
&x_i\sim\kappa(\omega_{z_i}),~\omega_k\sim g_0,~ \zv \sim\mbox{gCRP}(m,h_0,a,1-p), ~m\sim{\mbox{gNB}}\big(\big(\frac{p}{1-p}\big)^ah_0,a,p\big), % h_0p^a(1-p)^{-a},a,p), % \Big(\frac{p}{1-p}\Big)^a\Big),
 \eeqs
%The prediction rule of the reparameterized model is 
whose prediction rule is 
 \beq\label{eq:PredictRule1}
%p(z_{m+1} = k|\zv,m,\gamma_0,a,p)
P(z_{i} = k|\zv^{-i},m,h_0,a,p)  \propto
\begin{cases}
n_k^{-i} -a, &  {\mbox{for }} k=1,\cdots,l^{-i};\\ %\frac{n_k-a}{m+\gamma_0p^{-a}-al}, 
h_0 (1-p)^{-a}, & {\mbox{if } }k=l^{-i}+1. %\frac{\gamma_0 p^{-a}} {m+\gamma_0p^{-a}-al}, 
\end{cases}
\eeq
For the reparameterized gNBP, the number of clusters $l$ in a sample of size $m$ has PMF %is distributed as
\begin{align}\label{eq:f_L3}
f_L(l|m,h_0,a,p)  &=\frac{h_0^l (1-p)^{-al} S_a(m,l)}{\sum_{l=0}^m h_0^l (1-p)^{-al} S_a(m,l) },~l=0,\cdots,m.
\end{align}

Although $\gamma_0 p^{-a}\equiv h_0 (1-p)^{-a}$ and neither $\gamma_0$ nor $h_0$ directly influences the cluster sizes  $n_k\sim\mbox{TNB}(a,p)$, % distributions, 
below we show that whether $\gamma_0$ or $h_0$ is regularized with a prior distribution substantially impacts %the  distribution and 
the asymptotic behaviors of both the number and sizes of clusters.

\subsubsection{Connections to Normalized Generalized Gamma Process}
In a normalized generalized gamma process, it is standard to fix the scale parameter as one. We let $G=H\sim\mbox{gGaP}(H_0,a,1)$ in (\ref{eq:fxzv_G_N}), where $0\le a <1$ is required to ensure  $G(\Omega)>0$. % to be strictly positive. 
With $h_0:=H_0(\Omega)$ and $\beta: = \frac{p}{1-p}$, we can express (\ref{eq:betaEPPF}), the joint marginal distribution of $\zv$ and the auxiliary variable  $\beta$,  in terms of the ECPF of the reparameterized gNBP in (\ref{eq:f_Z_M2}) as
%we marginalize out $H$ of $f(\zv,\beta|H,m)$ in (\ref{eq:fxzv_G_N}), leading to a joint marginal distribution of the auxiliary variable  $\beta$ and $\zv$ as
\beqs \label{eq:f_Z_M1}
f\left(\zv,\beta\Big|m,h_0,\rho(dr)=\frac{r^{-a-1}}{\Gamma(1-a)}e^{-r}dr\right) %&= \E_{G}[ f(\zv,m|G) ]%= %\frac{1}{N!}\E\left[e^{-G(\Omega\backslash\mathcal{D}) }\right]  \prod_{k:\omega_k\in\mathcal{D}} \gamma_0 \int r_k^{n_k} e^{-r_k} \rho(dr)\notag\\
%=\frac{1}{m!}\E\left[e^{-G(\Omega\backslash\mathcal{D}) }\right]  \prod_{k:\omega_k\in\mathcal{D}} \gamma_0 \int r_k^{n_k} e^{-r_k} \rho(dr)\notag\\
= m\beta^{-1} %p^{-1}(1-p) 
 f(\zv,m|h_0,a,p).
%\frac{p^{m}}{(m)!}e^{-\frac{h_0(1-(1-p)^a-1)}{a(1-p)^a}}
% h_0^{l}  (1-p)^{al}
%\prod_{k=1}^{l{}} \frac{\Gamma(n_k-a)}{{\Gamma(1-a)} }
\eeqs 
As in \citet{lijoi2007controlling}, the EPPF $f(\zv|m,h_0,a)=\int_0^\infty f\big(\zv,\beta\big|m,h_0,\rho(dr)=\frac{r^{-a-1}}{\Gamma(1-a)}e^{-r}dr\big) d\beta$ is %can be expressed as
%and the EPPF as $f(\zv|m,a,h_0)=\int_0^\infty f\big(\zv,\beta\big|m,h_0,\rho(dr)=\frac{r^{-a-1}}{\Gamma(1-a)}e^{-r}dr\big) d\beta$, which is 
\beq
f(\zv|m,h_0,a)= \frac{1}{\Gamma(m)} {a^{l-1}e^{\frac{h_0}{a}} \sum_{i=0}^{m-1} \binom{m-1}{i}(-1)^i\left(\frac{h_0}{a}\right)^{i/a} \Gamma\left(l-\frac{i}{a},\frac{h_0}{a}\right)  \prod_{k=1}^l \frac{\Gamma(n_k-a)}{\Gamma(1-a)}},\notag
\eeq
where $\Gamma(a,x)=\int_x^\infty t^{a-1}e^{-t}dt$ is the incomplete gamma function. % \citep{lijoi2007controlling}.

The EPPF $f(\zv|m,h_0,a)$ of the  NRMI mixture %normalized generalized gamma process 
is consistent \citep{pitman2003poisson,csp,lijoi2007controlling} and $\beta$ is an auxiliary variable that can be neither  regularized nor fixed, whereas the EPPF of the (reparameterized) gNBP is generally an inconsistent EPPF that violates the addition rule and $\beta$ is a model parameter that can be either fixed or regularized with a prior distribution. 
%Although the marginal likelihoods (\ref{eq:f_Z_M1}) and (\ref{eq:f_Z_M2}) are very similar to each other, the EPPF $f(\zv|m,a,h_0)$ of the normalized generalized gamma process  mixture is distinct from the EPPF $f(\zv|m,a,h_0)$ of the (reparametrized) gNBP. 
Despite the distinctions on the interpretations of $\beta$, % and the EPPFs, % between the normalized generalized gamma process and gNBP, 
given the similarity between the likelihood functions (\ref{eq:f_Z_M1}) and (\ref{eq:f_Z_M2}), significant differences on posterior simulation and hence performance seem to be unlikely between the normalized generalized gamma process using the auxiliary variable  $\beta$ and the reparameterized gNBP. In fact, we observe similar clustering results %performance between posterior simulations 
using (\ref{eq:f_Z_M1}) and (\ref{eq:f_Z_M2})  for $0\le a <1$ and $0<p<1$. 
However, there are clear  differences, in both theory and experiments,  on the asymptotic behaviors of both the number and  sizes of clusters between the gNBP using (\ref{eq:f_Z_M}) and reparameterized gNBP using (\ref{eq:f_Z_M2}), as discussed below. %, %, even though they come from exactly the same model. % although with different representations. 
%the key difference between which is whether $\gamma_0$ or $h_0=\gamma_0 (\frac{1-p}{p})^a$ is treated as the mass parameter to be inferred, as discussed below. 

  \subsection{Asymptotics for the Number and Sizes of Clusters} % and Cluster Sizes} 
  \label{sec:asym}% for the generalized negative binomial process}

In a normalized generalized gamma process mixture model, \citet{lijoi2007controlling} showed that with a positive discount parameter $0<a<1$, the number of clusters increases at a power-law as a function of the sample size, and with a discount parameter $a$ close to one, there is a \emph{reinforcement} mechanism to favor a few large-size clusters in the posterior.  Although intriguing, there is lack of precise description about the distribution of the cluster sizes; in addition, $a<0$ is not allowed.  
Given similar likelihoods %\beqs \label{eq:f_Z_M1}
$f(\zv,\beta|m,a,h_0) %&= \E_{G}[ f(\zv,m|G) ]%= %\frac{1}{N!}\E\left[e^{-G(\Omega\backslash\mathcal{D}) }\right]  \prod_{k:\omega_k\in\mathcal{D}} \gamma_0 \int r_k^{n_k} e^{-r_k} \rho(dr)\notag\\
%=\frac{1}{m!}\E\left[e^{-G(\Omega\backslash\mathcal{D}) }\right]  \prod_{k:\omega_k\in\mathcal{D}} \gamma_0 \int r_k^{n_k} e^{-r_k} \rho(dr)\notag\\
= m\beta^{-1} %p^{-1}(1-p) 
 f(\zv|m,h_0,a,p) 
$ as in (\ref{eq:f_Z_M1}), we expect in the prior for $0\le a<1$, the distributions and  asymptotic behaviors of the cluster number and sizes to be similar between the normalized generalized gamma process mixture model using the auxiliary variable and the reparameterized gNBP count-mixture model. Thus the analysis of the reparameterized gNBP could also help better understand how the cluster number and sizes are controlled by model parameters in the normalized generalized gamma process.

For both the gNBP and reparameterized gNBP, we can precisely describe the cluster-size distributions as $n_k\sim\mbox{TNB}(a,p)$. 
However, the gNBP will be shown to have asymptotic behaviors on both the number and average size of clusters that are completely opposite to that of the reparameterized gNBP.
Recall that  $m\sim\mbox{gNB}(\gamma_0,a,p)$ is used to model the sample size, we have $\E[m]=%\E[X(\Omega)]=
\gamma_0\big(\frac{p}{1-p}\big)^{1-a} = h_0
\frac{p}{1-p}$. If $\gamma_0=G_0(\Omega)$ is assumed to be finite, then the increase of $m$ towards infinity would drive $p$ towards one and hence $h_0 = \gamma_0 \big(\frac{1-p}{p}\big)^a$ towards zero for $0<a<1$ and towards infinity for $a<0$, making $H\sim\mbox{gGaP}(H_0,a,1)$ not well defined. Whereas if $h_0=H_0(\Omega)$ is assumed to be finite, then the increase of $m$ towards infinity would drive  %drive $p$ towards one and hence 
$\gamma_0 = h_0 \big(\frac{p}{1-p}\big)^a$ towards infinity for $0<a<1$ and towards zero for $a<0$, making $G\sim\mbox{gGaP}(G_0,a,(1-p)/p)$ not well defined. Therefore, the gNBP  from (\ref{eq:gGaPP0}) and reparameterized gNBP from (\ref{eq:gGaPP}), despite their close connections, shall be treated differently when $a\neq 0$.

\subsubsection{Number of Clusters}\label{sec:clusterNumber}

\begin{figure}[!tb]
\begin{center}
\includegraphics[width=0.83\columnwidth]{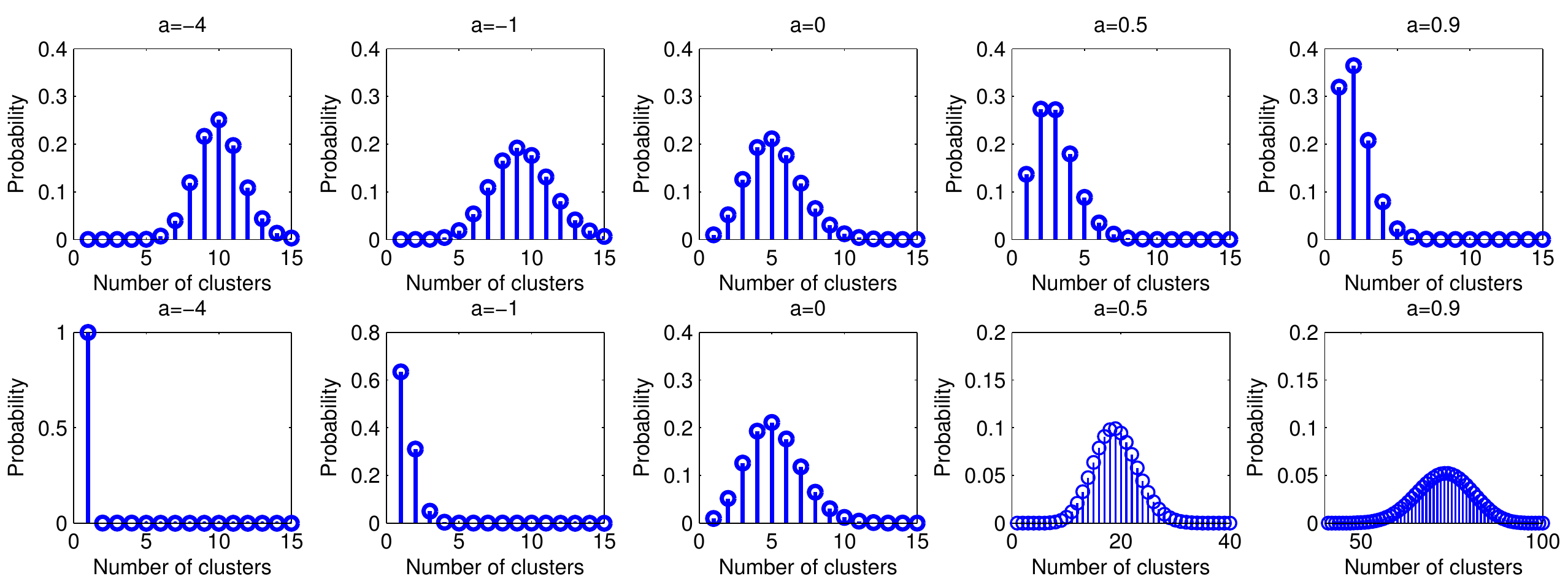}
\end{center}
\vspace{-4.0mm}
\caption{ \label{fig:PMF_CRT} \small
The \emph{a priori} probability mass functions (PMFs) of the number of clusters $l$ in a sample of size $m=100$ for $a=-4$, $-1$, $0$, $0.5$ and $0.9$. The first and second rows show the results for the gNBP and reparameterized gNBP, respectively. %The first to fifth columns show the results for $a=-4$, $-1$, $0$, $0.5$ and $0.9$, respectively. 
The mass parameters are set as $\gamma_0=1$ for the gNBP and $h_0=1$ for the reparameterized gNBP. The probability parameters $p$ for the gNBP and the reparameterized gNBP are calculated based on 
$m=\gamma_0\big(\frac{p}{1-p}\big)^{1-a}$ and $m=h_0\frac{p}{1-p}$, respectively. %Similar trends in the changes of the high density regions of the PMFs are observed by fixing $p$ at a constant.
\vspace{-2mm}
}
\end{figure}

To visualize the differences  between the cluster-number distributions, we show in Figure \ref{fig:PMF_CRT} the PMF $f_L(l|m,\gamma_0,a,p)$ of the gNBP as in (\ref{eq:f_L2}) and the PMF $f_L(l|m,h_0,a,p)$ of the reparameterized gNBP as in (\ref{eq:f_L3}), assuming $\gamma_0=1$, $h_0=1$ and a finite sample size $m=100$, where $m=\gamma_0\big(\frac{p}{1-p}\big)^{1-a}$ and $m=h_0\frac{p}{1-p}$ are used to calculate the probability parameter $p$ in the gNBP and reparameterized gNBP, respectively. It is evident that if the models behave in the way we expect in the prior, for the gNBP, the increase of $a$ would drive the high density region of the PMF towards left, encouraging fewer clusters, whereas for the reparameterized gNBP, the increase of $a$ would drive the high density region of the PMF towards right, encouraging more clusters.

  For the gNBP, with %(\ref{eq:GNBPdraw})
  (\ref{eq:ClusterStructure}), the expected number of  clusters and size of each cluster % distinct atoms, samples on each distinct atom,
  %and sample size 
  are $\E[l]=\gamma_0\frac{(1-(1-p)^a)}{ap^a}$ and $\E[n_k] = \frac{a(1-p)^{a}}{1-(1-p)^{a}}\frac{p}{1-p}$, % and $\E[m]=\E[X(\Omega)]=\gamma_0\big(\frac{p}{1-p}\big)^{1-a}$, 
  respectively. With $\E[m]=\E[l]\E[n_k]=\gamma_0\big(\frac{p}{1-p}\big)^{1-a}$, we have $p=1-(1+(\frac{\E[m]}{\gamma_0})^{\frac{1}{1-a}})^{-1}$ and hence for $a\neq 0$, we have
$
\E[l]= %\sim\mbox{Pois}\left(
\frac{\gamma_0}{a}\Big( {\big( \big(\frac{\gamma_0}{\E[m]}\big)^{\frac{1}{1-a}} + 1\big)^{a} -  {\big(\frac{\gamma_0}{\E[m]}\big)^{\frac{a}{1-a}}}  } \Big).\notag
%\right).
$
  When $a=0$, we have $\E[l] = -\gamma_0\ln(1-p)=\gamma_0(\ln({\E[m]+\gamma_0})-\ln\gamma_0)$. 
  Thus as $\E[m]$ % \rightarrow \infty$, 
  increases towards infinity,   if the model behaves in the way we expect,  we have
  \beq\label{eq:Pn_k1}
   \E[l]\sim
    \begin{cases}
\frac{\gamma_0^{\frac{1}{1-a}}}{-a}(\E[m])^{\frac{-a}{1-a}}, & \mbox{if } a< 0;\\
{\gamma_0}\ln(\E[m]), & \mbox{if } a= 0;\\
{\gamma_0}/{a}, & \mbox{if } 0< a<1.
%\frac{\gamma_0}{-a}(\frac{\E[m]}{\gamma_0})^{\frac{-a}{1-a}}, & \mbox{if } a< 0.\\
\end{cases}
  \eeq
  %as summarized in Table \ref{tab:cluster_number}, 
  For the gNBP count-mixture model,  the larger the discount parameter $a$ is, the more slowly the  number of clusters increases: %As shown in Figure \ref{fig:Asym},  %the expected number of clusters
  $\E[l]$ increases at a power-law rate for $a<0$, at a logarithmic rate for $a=0$, and towards a fixed value for $0<a<1$. The mass parameter $\gamma_0$  influences the asymptotic behaviors only with %linear 
  scaling.

%With (\ref{eq:GNBPdraw}),
For the reparameterized gNBP, substituting $\gamma_0$ with $h_0\big(\frac{p}{1-p}\big)^a$, % into the above analysis,  %the expected values of these three counts become %the expected number of  clusters, size of each cluster % distinct atoms, samples on each distinct atom,
  %and sample size become 
  we have
  $\E[l]=h_0\frac{(1-(1-p)^a)}{a(1-p)^a}$, $\E[n_k] = \frac{a(1-p)^{a}}{1-(1-p)^{a}}\frac{p}{1-p}$ %. % and $\E[m]=h_0\big(\frac{p}{1-p}\big)$, respectively. 
  and $\E[m]={h}_0\big(\frac{p}{1-p}\big)$. With $p=\frac{\E[m]}{h_0+\E[m]}$, for $a\neq 0$, we have % and hence
$
\E[l]= %\sim\mbox{Pois}\left(
\frac{h_0}{a}\Big( \Big(\frac{h_0+\E[m]}{h_0}\Big)^a - 1  \Big).\notag
%\right).
$ % for $a\neq 0$.
  When $a=0$, we have $\E[l] =-h_0\ln(1-p)=h_0(\ln(\E[m]+{h_0})-\ln(h_0)$. 
  Thus as $\E[m]$ increases towards infinity, if the model behaves in the way we expect, %\rightarrow \infty$, 
  we have
  \beq\label{eq:Pn_k1}
   \E[l]\sim
    \begin{cases}
    {h}_0/(-a)   , & \mbox{if } a< 0;\\
{h_0}\ln(\E[m]), & \mbox{if } a= 0;\\
\frac{h_0^{1-a}}{a} ({\E[m]})^a  , & \mbox{if } 0< a<1.
\end{cases}
  \eeq  
   %As summarized in Table \ref{tab:cluster_number}, 
   Apparently, reparameterizing  (\ref{eq:gGaPP0}) as (\ref{eq:gGaPP}) %$X\sim\mbox{PP}(G),~G\sim\mbox{gGaP}(G_0,a,(1-p)/p)$ as $X\sim\mbox{PP}(\frac{p}{1-p}H),~H\sim\mbox{gGaP}(H_0,a,1)$ %in the gNBP %makes the reparameterized gNBP have 
   leads to completely opposite asymptotic behaviors  that the larger the discount parameter $a$ is, the more rapidly the  number of clusters increases:  
   $\E[l]$ increases towards a fixed value for $a<0$, at a logarithmic rate for $a=0$, and at a power-law rate for $0<a<1$. %The mass parameter $h_0$  influences the asymptotic behavior also only with %linear 
 % scaling.  
Given the similarity between the likelihoods, it is unsurprising %to find 
that  when $0\le a <1$, the reparameterized gNBP count-mixture model has similar asymptotic behaviors as  the normalized generalized gamma process mixture model, whose prior number of clusters is shown in \citet{lijoi2007controlling} to increase as a power function of the sample size %exhibits power-law behavior 
for $0<a<1$. % and a logarithmic function for $a=0$.

\subsubsection{Sizes of Clusters}

\begin{figure}[!tb]
\begin{center}
\includegraphics[width=0.66\columnwidth]{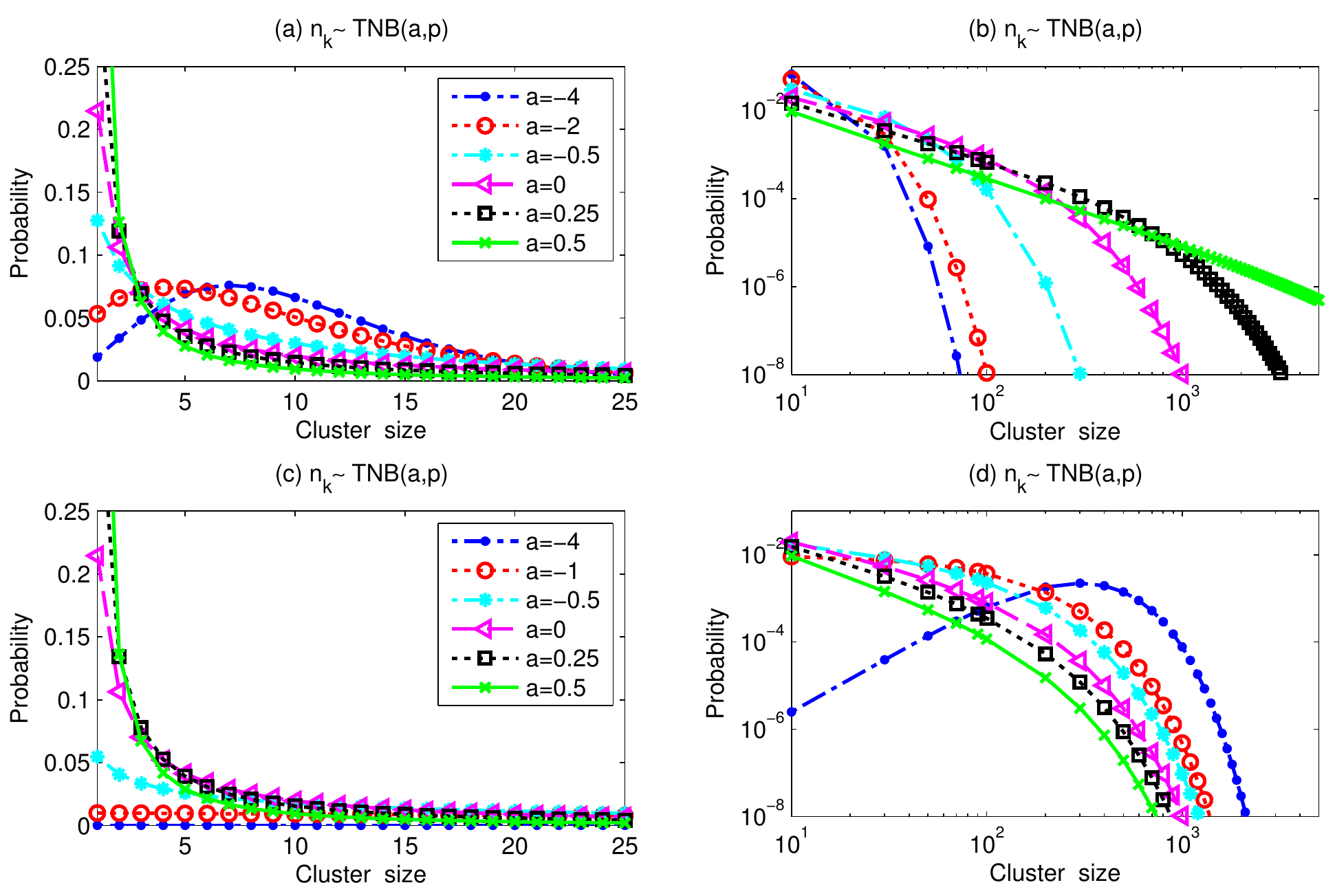}% GNBPdraw7.pdf}
\end{center}
\vspace{-6mm}
\caption{ \label{fig:GNBPdraw}  \small
The $a~priori$ PMFs of the cluster sizes $n_k\sim\mbox{TNB}(a,p)$ for $a=-4$, $-2$, $0$, $0.25$ and $0.5$. The first and second rows show the results of the gNBP and reparameterized gNBP, respectively. For the gNBP, the probability parameter $p$ is calculated under the assumption that $\E[m]= \gamma_0\big(\frac{p}{1-p}\big)^{1-a}=100$, whereas for the reparameterized gNBP, $p$ is calculated based on $\E[m]=h_0\frac{p}{1-p}=100$. We set $\gamma_0=1$ for the gNBP and $h_0=1$ for the reparameterized gNBP. 
\vspace{-2mm}
}
\end{figure}

The $a~priori$ cluster sizes %in both the gNBP and its reparameterized version 
follow $n_k\stackrel{iid}{\sim}\mbox{TNB}(a,p)$, whose PMF, for any $a> 1-{2}p^{-1}$ (hence any $a>-1$), always has the mode at $n_k=1$ and monotonically decreases.
In a sample of finite size, for both the gNBP and reparameterized gNBP, as $a$ increases towards one,  Figures \ref{fig:GNBPdraw} (a) and (c) show that the PMF of $n_k\sim\mbox{TNB}(a,p)$ gets a more peaked mode towards left. %which is always at $n_k=1$ if $a>1-2/p$ (note that $-1>1-2/p$),  thus 
%and hence the ratio of unit-size clusters would tend to increase. 
Figure \ref{fig:GNBPdraw} (b) shows that the PMF of $n_k\sim\mbox{TNB}(a,p)$ in the gNBP gets heavier tails (decays more slowly) as $a$ increases towards one, whereas Figure \ref{fig:GNBPdraw} (d) shows that the tails decay rapidly at similar rates % behaviors are similar 
for different $a$ in the repararemterized gNBP.

   Using the PMF of $n_k\sim\mbox{TNB}(a,p)$ in (\ref{eq:TNB}), 
  the probability for $n_k=j$, $j=1,2,\cdots,$ is 
 $
  P(n_k=j|a,p)%= \frac{\Gamma(j-a)}{j!\Gamma(-a)}\frac{p^j(1-p)^{-a}}{1-(1-p)^{-a}} 
  = \frac{\Gamma(j-a)}{j!\Gamma(1-a)}p^{j-1}\frac{ap}{1-(1-p)^{a}},
  $
  where the term $\frac{ap}{1-(1-p)^{a}}$ monotonically decreases towards $\max\{a,0\}$ as $p$ increases towards one. Thus as $\E[m]\rightarrow \infty$, we have $p\rightarrow 1$ and hence %$P(n_k=1|a,p) \ge \max\{a,0\}$ and 
  \beq
  P(n_k=j|a,p) \sim \frac{a\Gamma(j-a)}{j!\Gamma(1-a)} p^{j-1}, ~~\mbox{if } 0<a<1, %\max\{a,0\}.
  \eeq
  which are similar to the properties of the normalized generalized gamma process analyzed in \citet{lijoi2007controlling}.
The discount parameter determines the lower bounds of the \emph{a priori} ratio of unit-size clusters (clusters with only one sample)
as
\beq
P(n_k=1|a,p) \ge \max\{a,0\}.
\eeq
 %, which are equal to $\max\{a,0\}$. % that would have only one sample, % \emph{a priori}, which
%which in the prior decreases towards $\max\{a,0\}$. % as $p$ increases towards one.
E.g., if $a=0.90$, then in the prior at least $90\%$ of the instantiated clusters would be unit size. 
%For $0<a<1$, the asymptotic behavior on the cluster sizes is the same for the gNBP, the reparameterized gNBP and the normalized generalized gamma process in \citet{lijoi2007controlling}.

\begin{table}[!tb]
\caption{\small Asymptotic behaviors of the average size  of clusters.}\label{tab:cluster_size}
\begin{center}\small 
\begin{tabular}{|c|c|c|c|}
\hline
Model & $a<0$ & $a=0$ & $0<a<1$ \\
\hline
gNBP, $\E[n_k] $& $\frac{-a}{\gamma_0^{\frac{1}{1-a}}} (\E[m])^{1-\frac{-a}{1-a}}$  & $\frac{1}{\gamma_0}\frac{\E[m]}{\ln(\E[m])}$ & $\frac{a}{\gamma_0} \E[m]$\\ 
\hline
Reparameterized gNBP,  $\E[n_k] $& $\frac{-a}{h_0}\E[m]$& $\frac{1}{h_0} \frac{\E[m]}{\ln(\E[m])}$ &  $\frac{a}{h_0^{1-a}} (\E[m])^{1-a}$\\ 
\hline
\end{tabular}
\end{center}
\vspace{-3mm}
\end{table}%

Similar to the analysis in Section \ref{sec:clusterNumber}, we summarize in Table \ref{tab:cluster_size} the asymptotic behaviors of the average size of clusters  $\E[n_k]$ for both the gNBP and reparameterized gNBP.
 It is clear that, asymptotically, as the discount parameter $a$ increases, the average size of clusters increases at a higher rate  for the gNBP while at a lower rate %the cluster-size average increases 
  for the reparameterized gNBP.

\subsubsection{Controlling the Asymptotic Behaviors} %Model and Parameter Selection}

In summary, for the gNBP, increasing $a$ towards one %, the number of clusters tends to decrease but the ratio of unit-size clusters tends to increase, encouraging 
would encourage a smaller number of larger clusters together with a higher ratio of unit-size clusters to fit the data; whereas decreasing $a$ increases towards $-\infty$ %, the number of clusters tends to increase and the cluster sizes tend to concentrate more around its mean, %be less overdispersed, 
would encourage a larger number of similar-size smaller clusters. For the reparameterized gNBP, increasing $a$ towards one %, both the number of clusters and the ratio of unit-size clusters tend to increase, encouraging 
would encourage a larger number of smaller clusters with a higher ratio of unit-size clusters; whereas decreasing  $a$ towards $-\infty$ %, the number of clusters tends to decrease and the cluster sizes tend to be less overdispersed, encouraging 
would encourage a smaller number of similar-size larger clusters. When $a=0$, both the gNBP and reparameterized gNBP become the NB process in \citet{NBP2012}, which is closely related to the CRP in that they have the same EPPF and prediction rule. The NB process has the advantages over the CRP that given the probability parameter $p$, which has an analytic beta conditional posterior, the conditional posterior of the mass parameter $\gamma_0$ follows an analytic gamma distribution and the $a~priori$ cluster sizes are known to follow the $\mbox{Log}(p)$ distribution. In the CRP, the concentration parameter $\gamma_0$ is usually sampled with a data augmentation approach of \citet{Escobar1995}.

\subsection{Size-Dependent Exchangeable Partition Probability Functions}

With $z_{1:j} := (z_1,\cdots,z_j)$ and $l_{(j)}$ representing the number of clusters in $z_{1:j}$, we have $\gamma_0\int_0^\infty re^{-r}\rho(dr)+ \sum_{k=1}^{l_{(j)}}\frac{\int_{0}^\infty r^{n_k+1} e^{-r} \rho(dr)}{\int_{0}^\infty r^{n_k} e^{-r} \rho(dr)} = p(\gamma_0p^{-a}+ j - a l_{(j)})$ for the gNBP. Using Corollary \ref{sdEPPF}, we have
\small
\beqs
{f(z_{1:m-1}|m,\gamma_0,a,p)}%&=\frac{j!f(z_{1:j}|j,\gamma_0,\rho)f_J(j|\gamma_0,\rho)}{m!f_M(m|\gamma_0,\rho)} p^{m-j} \prod_{i=j}^{m-1}(\gamma_0p^{-a}+ i- a l_{(i)})\\
=f(z_{1:m-1}|m-1,\gamma_0,a,p)\frac{\sum_{l=0}^{m-1} \gamma_0^lp^{-al}S_a(m-1,l)}{\sum_{l=0}^m \gamma_0^lp^{-al}S_a(m,l)}\left(\gamma_0p^{-a}+ (m-1)- a l_{(m-1)}\right),
\eeqs
\normalsize
and we may also marginalize out $z_{m-1}$ as
$
{f(z_{1:m-2}|m,\gamma_0,\rho)}%&=\frac{j!f(z_{1:j}|j,\gamma_0,\rho)f_J(j|\gamma_0,\rho)}{m!f_M(m|\gamma_0,\rho)} p^{m-j} \prod_{i=j}^{m-1}(\gamma_0p^{-a}+ i- a l_{(i)})\\
={f(z_{1:m-2}|m-2,\gamma_0,\rho)}\frac{\sum_{l=0}^{m-2} \gamma_0^lp^{-al}S_a(m-2,l)}{\sum_{l=0}^m \gamma_0^lp^{-al}S_a(m,l)}$ 
%$\left(\gamma_0p^{-a}+ (m-2)- a l_{(m-2)}\right)$ 
$ \left[( \gamma_0p^{-a}+ (m-1)- a l_{(m-2)})(m-2-al_{(m-2)}) +( \gamma_0p^{-a}+ (m-1)- a (l_{(m-2)}+1))\gamma_0p^{-a}  \right].
$ Similar analysis can be further carried out to marginalize out $z_{m-3},z_{m-4},\cdots$. % but the expressions of the marginal probability functions quickly become rather complicated.
It is easy to verify that $f(z_{1:j}|m,\gamma_0,\rho)\equiv f(z_{1:j}|j,\gamma_0,\rho)$ for $m\ge j$ when $a=0$. 
When $a\neq 0$, $f(z_{1:j}|m,\gamma_0,\rho)$ is usually not equal to  $f(z_{1:j}|j,\gamma_0,\rho)$ for $m>j$. E.g.,  since
\begin{align}
f(z_{1:2}|3,\gamma_0,\rho)
%&=f(z_{1:2}|2,\gamma_0,\rho)\frac{\sum_{l=0}^2 \gamma_0^lp^{-al}S_a(2,l)}{\sum_{l=0}^3 \gamma_0^lp^{-al}S_a(3,l)} (\gamma_0p^{-a}+ 2- l_{(2)} a)\\
&=f(z_{1:2}|2,\gamma_0,\rho) \frac{(1-a+\gamma_0p^{-a})(\gamma_0p^{-a}+2-l_{(2)}a)}{(1-a)(2-a)+(3-3a)\gamma_0p^{-a}+\gamma_0^2p^{-2a}},
\end{align}
if $a\neq 0$, then $f(z_{1:2}|3,\gamma_0,\rho)\neq f(z_{1:2}|2,\gamma_0,\rho)$ regardless of whether $l_{(2)}=1$ or $l_{(2)}=2$. The reparameterized gNBP can be similarly analyzed by replacing $\gamma_0$ with $h_0\big(\frac{p}{1-p}\big)^a$.

\begin{figure}[!tb]
\begin{center}
\includegraphics[width=0.9\columnwidth]{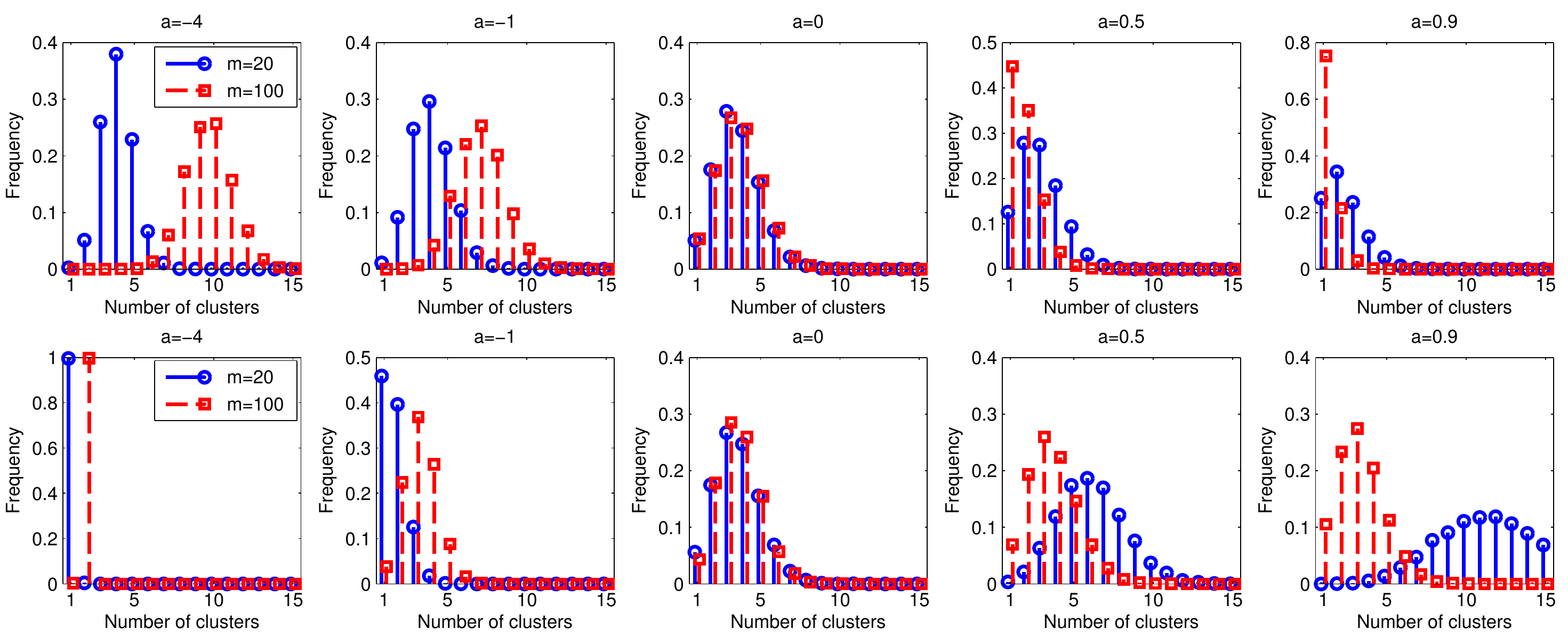}
\end{center}
\vspace{-6mm}
\caption{ \label{fig:sdEPPF}  \small
Comparison of the \emph{a priori} distributions of the number of clusters in $z_{1:20}$ between  a sample of size $m=20$ and a sample of size $m=100$. The first and second rows show the results for the gNBP, simulated with $\gamma_0=1$ and $p=0.9$, and reparameterized gNBP,  simulated with $h_0=1$ and $p=0.9$, respectively. Columns from left to right are the results for $a=-4$, $-1$, $0$, $0.5$ and $0.9$, respectively. 
\vspace{-0mm}
}
\end{figure}

In a cluster structure, the sample size $m$ usually plays an important role on how a subset of size $j$ is clustered. As shown in the first row of Figure \ref{fig:sdEPPF}, we compare  the simulated \emph{a priori} cluster-number distributions $f(l_{(20)}|m=20,\gamma_0,a,p)$ and $f(l_{(20)}|m=100, \gamma_0,a,p)$ for the gNBP, where $\gamma_0=1$ and $p=0.9$; %, where $l_{(20)}$ represents the number of clusters in $z_{1:20}=(z_1,\cdots,z_{20})$; 
columns from left to right are the results for  $a=-4$, $-1$, $0$, $0.5$ or $0.9$, respectively.  %For $f(l_{(20)}|m=20, \gamma_0,a,p)$, we plot both its theoretic PMFs and the simulated normalized  histograms; for $f(l_{(20)}|m=100, \gamma_0,a,p)$, we plot the simulated normalized histograms. 
It is clear that except for $a=0$,  for which $P(\Pi_j|m)\equiv P(\Pi_m)$ for $m\ge j$, $f(l_{(20)}|m=20, \gamma_0,a,p)$ is obviously different from $f(l_{(20)}|m=100,  \gamma_0,a,p)$.
Similarly, in the second row of Figure \ref{fig:sdEPPF} we compare the cluster-size distributions $f(l_{(20)}|m=20,h_0,a,p)$ and $f(l_{(20)}|m=100, h_0,a,p)$ for the generalized gNBP, where $h_0=1$ %, $a=-4$, $-1$, $0$, $0.5$ or $0.9$, 
and $p=0.9$.   Except for $a=0$, these two cluster-size distributions are also obviously different.
 % normalized from 10,000 MCMC samples. We construct a Markov chain to draw exchangeable random partitions of $[20]$ based on the prediction rule and record the number of clusters in each MCMC sample. 
Note that for a consistent EPPF, we can simulation an exchangeable random partition by sequentially assigning the $i+1$ element to $\Pi_i$ using the prediction rule until the $m$th element is assigned \citep{csp}. For an inconsistent EPPF, as discussed in Section \ref{sec:thms}, %we need to construct a Markov chain to 
 we construct a Gibbs sampler based on the prediction rule to simulate exchangeable random partitions; 
for both $m=20$ and $m=100$, we consider 15000 Gibbs sampling iterations  and record the number of clusters in $z_{1:20}$ of the last 10,000 iterations, which are used to estimate the \emph{a priori} distributions %obtain the normalized histograms 
of the number of clusters. The sequential construction of inconsistent exchangeable random partitions for the generalized Chinese restaurant process is currently under investigation.

\section{MCMC  Inference}\label{sec:inference}

We  present %both a conditional sampler and 
Markov chain Monte Carlo (MCMC) inference for count-mixture modeling. Using the prediction rule  in Theorem \ref{thm:predict}, a generalized  P\'{o}lya urn sampling scheme is developed by generalizing the sampling algorithms developed for Dirichlet process mixture models \citep{Escobar1995,MullerDP,neal2000markov,GreenDP}. The parameters of the completely random measure $G$ can be inferred based on the ECPF in Theorem \ref{thm:ECPF}. 
We consider both the gNBP count-mixture model and its reparameterized version. We develop MCMC inference for both the gNBP based on the ECPF in (\ref{eq:f_Z_M}) and the prediction rule in (\ref{eq:PredictRule}), and the reparameterized gNBP based on the ECPF in (\ref{eq:f_Z_M2}) and prediction rule in (\ref{eq:PredictRule1}).%\\

\emph{\textbf{Sample $z_i$}}. The data samples are assumed independent conditioning on $G$ and hence exchangeable. % we can treat the $i$th data sample as the last one and sample its 
%We sample the cluster index $z_i$ conditioning on  $\zv^{-i}:=\zv \backslash z_i$. % (z_1,\cdots,z_{i-1},z_{i+1},z_m)$.
Using the prediction rule of the gNBP in (\ref{eq:PredictRule}),
%For the gNBP, using (\ref{eq:PredictRule}), 
since $P(z_i=k|\zv^{-i},m,x_i,\mathcal{D}_m^{-i})\propto \kappa(x_i|\omega_k)P(z_i=k|\zv^{-i},m)$, where $\mathcal{D}_m^{-i}=\{\omega_k\}_{k:n_k^{-i}>0}$ and $n_k^{-i}=\sum_{j\neq i}\delta(z_j=k)$, we have%
%(z_1,\cdots,z_{i-1},z_{i+1},\cdots,z_m)^T$,
%we have $P(z_{i} = k|\zv^{-i},x_i,\mathcal{D}_m^{-i})  \propto (n_k^{-i}-a)\kappa(x_i|\omega_{k}),  \mbox{for }  \omega_k\in\mathcal{D}_m^{-i}$ and $P(z_{i} = k|\zv^{-i},x_i,\mathcal{D}_m^{-i})  \propto \gamma_0 p^{-a} \int_{\Omega}\kappa(x_i|\omega)g_0(\omega)d\omega,  \mbox{if } \omega_k  \in\Omega\backslash\mathcal{D}_m^{-i}$%the conditional posterior of $z_i$ as
%\beq\label{eq:zpost}
%%p(z_{m+1} = k|\zv,m,\gamma_0,a,p)
%P(z_{i} = k|\zv^{-i},\xv)  \propto
%\begin{cases}
%(n_k^{-i}-a)\int_{\Omega}\kappa(x_i|\omega_{k})g(\omega_k|\xv^{-i},\zv^{-i})d\omega_k, & \mbox{for }  \omega_k\in\mathcal{D}_m^{-i}; \\
%\gamma_0 p^{-a} \int_{\Omega}\kappa(x_i|\omega)g_0(\omega)d\omega, & \mbox{if } \omega_k  \in\Omega\backslash\mathcal{D}_m^{-i};
%\end{cases}
%\eeq
\beq\label{eq:zpost}
%p(z_{m+1} = k|\zv,m,\gamma_0,a,p)
P(z_{i} = k|\zv^{-i},m,x_i,\mathcal{D}_m^{-i},\gamma_0,a,p)  \propto
\begin{cases}
(n_k^{-i}-a)\kappa(x_i|\omega_{k}), & \mbox{for }  \omega_k\in\mathcal{D}_m^{-i}; \\
\gamma_0 p^{-a} \int_{\Omega}\kappa(x_i|\omega)g_0(\omega)d\omega, & \mbox{if } \omega_k  \in\Omega\backslash\mathcal{D}_m^{-i};
\end{cases}
\eeq
where $\xv^{-i}=\xv \backslash x_i$ %
%(x_1,\cdots,x_{i-1},x_{i+1},\cdots,x_m)^T$
and
%\beqs\label{eq:gomega_k}
%&g(\omega_k|\xv^{-i},\zv^{-i}) \propto { g_0(\omega)\prod_{j:(z_j=k,j\neq i)}\kappa(x_j|\omega_k)}. %{\int_{\Omega} g_0(\omega_k)\prod_{k:z_i=k}\kappa(x_i|\omega_k)d\omega_k}
%\eeqs %is the conditional posterior of the mixture component $\omega_k$.
$
P(\omega_k|\xv^{-i},\zv^{-i}) \propto { g_0(\omega)\prod_{j:(z_j=k,j\neq i)}\kappa(x_j|\omega_k)}. %{\int_{\Omega} g_0(\omega_k)\prod_{k:z_i=k}\kappa(x_i|\omega_k)d\omega_k}
$ %is the conditional posterior of the mixture component $\omega_k$.
As in \citep{neal2000markov}, we  further marginalize out $\omega_k\in\mathcal{D}_m^{-i}$ in (\ref{eq:zpost}), leading to 
\beq\label{eq:zpost0}
%p(z_{m+1} = k|\zv,m,\gamma_0,a,p)
P(z_{i} = k|\zv^{-i},m,\xv,\gamma_0,a,p)  \propto
\begin{cases}
(n_k^{-i}-a)\int_{\Omega}\kappa(x_i|\omega_{k})P(\omega_k|\xv^{-i},\zv^{-i})d\omega_k, & \mbox{for }  \omega_k\in\mathcal{D}_m^{-i}; \\
\gamma_0 p^{-a} \int_{\Omega}\kappa(x_i|\omega)g_0(\omega)d\omega, & \mbox{if } \omega_k  \in\Omega\backslash\mathcal{D}_m^{-i};
\end{cases}
\eeq
%$P(z_{i} = k|\zv^{-i},\xv)  \propto
%(n_k^{-i}-a)\int_{\Omega}\kappa(x_i|\omega_{k})P(\omega_k|\xv^{-i},\zv^{-i})d\omega_k$ for $\omega_k\in\mathcal{D}_m^{-i}$.
If %the mixture component prior
$g_0(\omega)$ is conjugate to  $\kappa(x|\omega)$ in this paper, the conditional posteriors $P(\omega_k|\xv^{-i},\zv^{-i})$ %in (\ref{eq:gomega_k})
and the integrals in (\ref{eq:zpost}) and (\ref{eq:zpost0}) can all be analytically calculated. For the reparameterized gNBP, we replace $\gamma_0p^{-a}$ with $h_0(1-p)^{-a}$ in both (\ref{eq:zpost}) and (\ref{eq:zpost0}).

\emph{\textbf{Sample $\omega_k$}}. The conditional posterior of an atom $\omega_k$ can be expressed as
\beq\label{eq:sample_omega_k}
P(\omega_k|-)\propto g_0(\omega)\prod_{i:z_i=k}\kappa(x_i|\omega_k) .
\eeq

\emph{\textbf{Sample $\gamma_0$ or $h_0$}}. For the gNBP, the ECPF in (\ref{eq:f_Z_M}) defines a likelihood for $\gamma_0$, $a$ and $p$. With a gamma prior $\mbox{Gamma}(e_0,1/f_0)$ placed on $\gamma_0$, %using the ECPF in  (\ref{eq:f_Z_M}), %(\ref{eq:PoisTNB}) and (\ref{eq:GNB_cor}), % (\ref{eq:Liklihood}) and (\ref{eq:Exptation0}),
we have %the conditional posterior of $\gamma_0$ as
\beqs
(\gamma_0|-)\sim\mbox{Gamma}\bigg(e_0 + l{},\frac{1}{f_0+ \frac{1-(1-p)^a}{ap^a}}\bigg).
\eeqs
%where  $L(\Omega) \ge |\mathcal{D}|$, with $L(\Omega) \equiv |\mathcal{D}|$ if and only if $G_0$ is a finite and continuous base measure.
As $a\rightarrow 0$, we have
$
\lim_{a\rightarrow 0}(\gamma_0|-)\sim\mbox{Gamma}\left(e_0 + l{},\frac{1}{f_0- \ln(1-p)}\right), %\notag
$
the same as that of the NB process count-mixture model \citep{NBP2012}. Similarly, for the reparameterized gNBP, with the ECPF in (\ref{eq:f_Z_M2}) and a gamma prior $\mbox{Gamma}(e_0,1/f_0)$ placed on $h_0$, we have 
$
(h_0|-)\sim\mbox{Gamma}\bigg(e_0 + l{},\frac{1}{f_0+ \frac{1-(1-p)^a}{a(1-p)^a}}\bigg)
$
and $
\lim_{a\rightarrow 0}(h_0|-)\sim\mbox{Gamma}\left(e_0 + l{},\frac{1}{f_0- \ln(1-p)}\right). %\notag
$

\textbf{\emph{Sample $a$}}.
Since $a<1$, we have $\tilde{a}=\frac{1}{1+(1-a)} \in(0,1)$. With a uniform prior placed on $\tilde{a}$ in $(0,1)$ and the likelihood of gNBP in (\ref{eq:f_Z_M}), we use the  griddy-Gibbs sampler \citep{griddygibbs} to
sample $a$ from a discrete distribution
\beq
(a|-)\propto f(\zv,m|\gamma_0,a,p)
\eeq
over a grid of points $\frac{1}{1+(1-a)}=0.0001,0.0002,\cdots,0.9999$. Similarly, with %the likelihood in 
 (\ref{eq:f_Z_M2}) for the reparameterized gNBP,  we sample $a$ from a grid of points with $
(a|-)\propto f(\zv,m|h_0,a,p)
$.

\emph{\textbf{Sample $p$}}. When $a\rightarrow 0$, the likelihood of the gNBP in (\ref{eq:f_Z_M}) becomes proportional to $p^{m}(1-p)^{\gamma_0}$, with a beta prior $\mbox{Beta}(a_0,b_0)$ on $p$, we have 
$
\lim_{a\rightarrow 0}(p|-)\sim\mbox{Beta}(a_0+m,b_0+\gamma_0),
$
 the same as that of the NB process count-mixture model \citep{NBP2012}.
When $a\neq 0$,  with %We %place
a uniform prior placed on $p$ in $(0,1)$ and the likelihood in (\ref{eq:f_Z_M}),  we use the  griddy-Gibbs sampler to sample $p$ from a discrete distribution
\beq
(p|-)\propto e^{-\frac{\gamma_0(1-(1-p)^a)}{ap^{a}}}
 p^{m-al{}}
\eeq over %consider
a grid of points $p=0.0001,0.0002,\cdots,0.9999$.
Similarly, for the reparameterized gNBP, for $a=0$, we have $
(p|-)\sim\mbox{Beta}(a_0+m,b_0+h_0),
$ and for $a\neq 0$, we sample $p$ from a grid of points using 
$
(p|-)\propto e^{-\frac{h_0(1-(1-p)^a)}{a(1-p)^{a}}}
 p^{m}(1-p)^{-al}. 
$ 

Note that for the normalized generalized gamma process mixture model \citep{lijoi2007controlling,james2009posterior}, since its likelihood with the auxiliary variable $\beta=\frac{p}{1-p}$ is related to the ECPF of the reparameterized gNBP as in (\ref{eq:f_Z_M1}), its inference closely follows that of the reparameterized gNBP, except that $a<0$ is not allowed and $p$ is an auxiliary variable that cannot be fixed and shall be sampled with
$
(p|-)\propto e^{-\frac{h_0(1-(1-p)^a)}{a(1-p)^{a}}}
 p^{m-1}(1-p)^{-al+1}
$ for $0<a<1$ and with $
(p|-)\propto 
 p^{m-1}(1-p)^{1+h_0}
$ for $a=0$. %Unsurprisingly, the differences between the likelihoods as in (\ref{eq:f_Z_M1}) 
We find that this difference in inference between the normalized generalized gamma process and  reparameterized gNBP is not significant enough to result in  major performance differences in our experiments for $0\le a<1$.

\section{Example Results}\label{sec:Results}

To cluster $m$ $P$ dimensional data vectors
$\{\xv_i\}_{1,m}$, %with $\kappa(\xv|\muv)=\mathcal{N}(\xv;\muv_k,\varphi^{-1})$ and $g_0(\muv_k)=\mathcal{N}(\muv_k; \mu_0,\varphi^{-1}_0\Imat_P)$, 
we construct a generalized negative binomial process (gNBP) Gaussian count-mixture model as
%\begin{align}
\beqs
&\xv_i\sim\mathcal{N}(\muv_{z_i},\varphi^{-1}\Imat_P),~\muv_k\sim\mathcal{N}(\muv_0,\varphi_0^{-1}\Imat_P),~\varphi\sim\mbox{Gamma}(c_0,1/d_0),\notag\\
&z_i\sim\sum_{k=1}^K \frac{r_{k}}{G(\Omega)} \delta_k, ~m\sim{\mbox{Pois}}(G(\Omega)),~G\sim{{}}{\mbox{gGaP}}(G_0,a,(1-p)/p),
\eeqs
where $K=\infty$ if $0\le a <1$ and $K\sim \mbox{Pois}(\frac{\gamma_0(1-p)^a}{-ap^a})$ if $a<0$. The reparameterized version of the model is constructed by replacing $m\sim{\mbox{Pois}}(G(\Omega)),~G\sim{{}}{\mbox{gGaP}}(G_0,a,(1-p)/p)$ with 
\beq
m\sim{\mbox{Pois}}\left(G(\Omega){p}/({1-p})\right),~G\sim{{}}{\mbox{gGaP}}(H_0,a,1).
\eeq Similar to the settings in \citet{ishwaran2000markov,griffin2010default}, we consider noninformative priors by letting $\muv_0\sim\mathcal{N}(0,1000\Imat_P)$, $\varphi_0\sim\mbox{Gamma}(0.001,1/0.001)$, and $c_0=d_0=0.001$. We set the gamma hyper-parameters for both $\gamma_0$ and $h_0$ as % let $\gamma_0=G_0(\Omega) \sim \mbox{Gamma}(e_0,1/f_0)$, with 
$e_0=f_0=1$. We place a noninformative beta prior $\mbox{Beta}(0.01,0.01)$ on $p$ when $a=0$, or a uniform prior on $p$ in $(0,1)$ when $a\neq 0$. Since $a<1$, we place a uniform prior on $\tilde{a}=\frac{1}{1+(1-a)}$ in $(0,1)$.

%where $h_0=H_0(\Omega) = \gamma_0 \big(\frac{1-p}{p}\big)^a$.
For the gNBP, 
using (\ref{eq:zpost0}), we have
\beq\label{eq:zpost2}
%p(z_{m+1} = k|\zv,m,\gamma_0,a,p)
P(z_{i} = k|\zv^{-i},m,\xv,\gamma_0,a,p)  \propto
\begin{cases}
(n_k^{-i}-a)\mathcal{N}\left(\xv_i; \muv_k^{-i},\varphi^{-1}\Imat_P+(\varphi_0+n_k\varphi)^{-1}\Imat_P\right), & \mbox{for }  \muv_k\in\mathcal{D}_m^{-i}; \\
\gamma_0 p^{-a}  \mathcal{N}\left(\xv_i; \muv_0,\varphi_0^{-1}\Imat_P+\varphi^{-1}\Imat_P\right), & \mbox{if } \muv_k  \in\Omega\backslash\mathcal{D}_m^{-i};\notag
\end{cases}
\eeq
where $\muv_k^{-i}:=\frac{\varphi_0\muv_0+\varphi\sum_{j: (z_j=k, j\neq i)}\xv_j}{\varphi_0+ n_k^{-i}\varphi}$, 
$(\varphi|-)\sim\mbox{Gamma}\left(c_0+\frac{mP}{2}, \frac{1}{d_0+\sum_k\sum_{i:z_i=k}\frac{\|\xv_i-\muv_k\|_2^2}{2}}\right)$, 
%\emph{\textbf{Sample $\mu_0$} and $\varphi_0$}. Using the conjugacy, we have
$(\muv_0|-)\sim\mathcal{N}\left(\frac{\varphi_0\sum_{k}\muv_k}{10^{-3}+l{}\varphi_0},\frac{1}{10^{-3}+l{} \varphi_0}\Imat_P \right)$ and $(\varphi_0|-)\sim\mbox{Gamma}\left(10^{-3}+\frac{lP}{2},\frac{1}{10^{-3}+\sum_{k}\frac{\|\muv_k-\muv_0\|_2^2}{2}}\right)$. The mass parameter $\gamma_0$, probability parameter $p$ and discount parameter $a$ are sampled as in Section \ref{sec:inference}. For the reparameterized gNBP Gaussian count-mixture model, the update equations can be similarly derived.

We consider the Galaxy dataset \citep{roeder1990density}, which consists of the relative velocities of $m=82$ galaxies. 
We consider 15,000 MCMC iterations, with the last 10,000 samples collected. The $m$ data points appear in a random order in each MCMC iteration. We let both $p$ and $\gamma_0$ be inferred from the data. %For initialization, we fix $a=0$ for the first 100 iterations  for all cases. 
We consider either fixing $a=-4$, $-2$, $-1$, $-0.50$, $0$, $0.25$, $0.50$, $0.90$ or $0.99$, or letting $a$ be inferred from the data. We record in each MCMC iteration the number of clusters, the ratio of unit-size clusters and the cluster-size distribution. %To better understand the asymptotic behavior of the number of clusters and the sizes of these clusters,
%We consider mixture modeling of both $\{x_{2i-1}\}_{1,41}$ and  $\{x_{i}\}_{1,82}$.

\begin{figure}[!tb]
\begin{center}
\includegraphics[width=0.76\columnwidth]{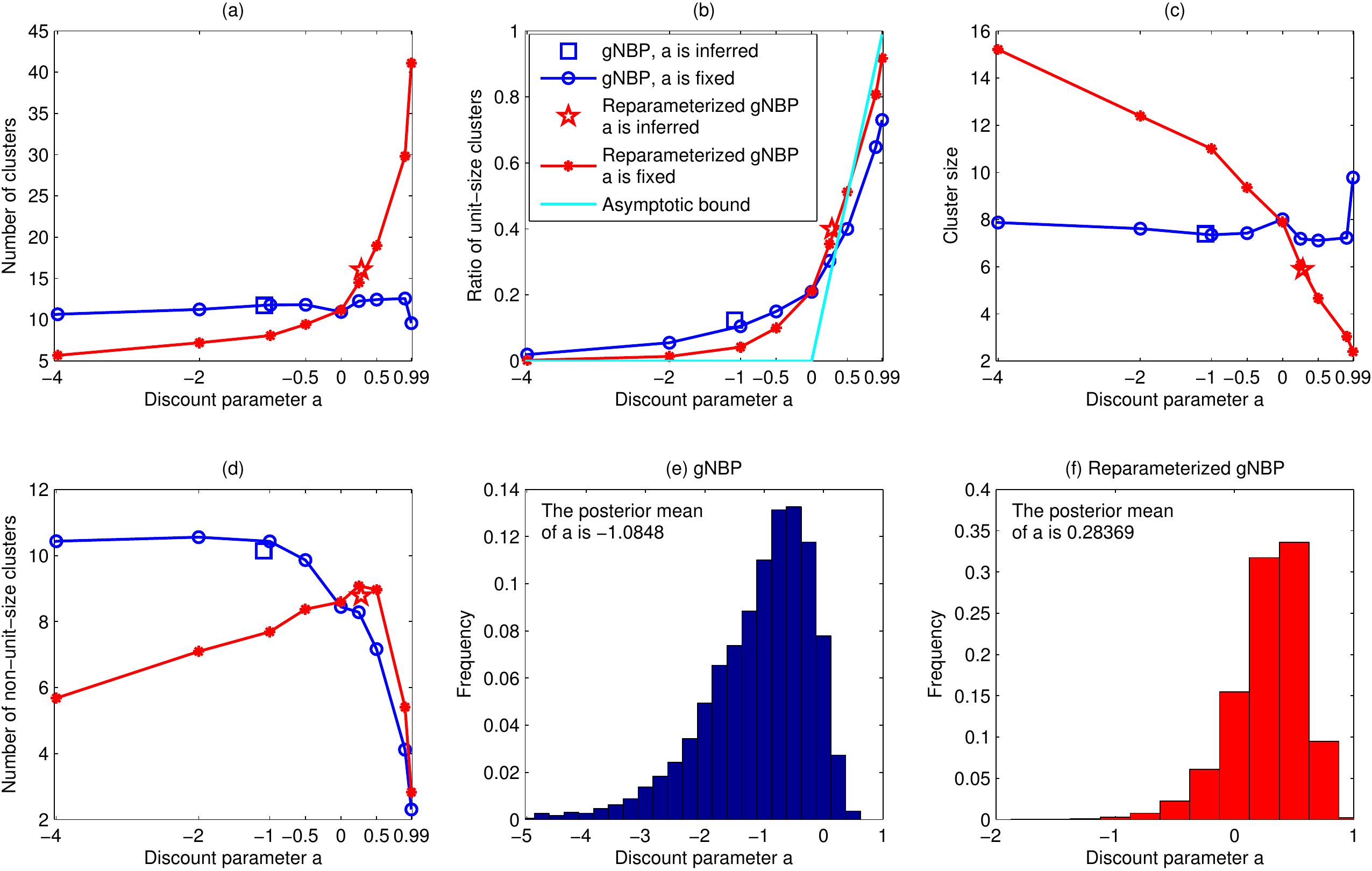}
\end{center}
\vspace{-5.5mm}
\caption{ \label{fig:clusters} \small
The posterior means of (a)   the number of clusters, (b) the ratio of unit-size clusters, (c) the average size of clusters and (d) the number of non-unit-size clusters, as a function of the discount parameter $a$, which is either fixed at $a=-4,-2,-0.50,0, 0.25,0.50,0.90,0.99$ or inferred from the data, for both the generalized negative binomial process (gNBP) count-mixture model and its reparameterized version. The normalized histograms of the inferred discount parameter $a$ for the gNBP and its reparameterized version are shown in (e) and (f), respectively.  
The prior asymptotic lower bound $\max\{a,0\}$ for the ratio of unit-size clusters is shown in (b).
%The normalized histograms of the collected MCMC samples of $a$ are shown in (c) for the gNBP and in (d) for the reparameterized gNBP.
The horizontal and vertical positions
of the $\Box$ sign for the gNBP and $\bigstar$ sign for the reparameterized gNBP %in both (a) and (b) 
are the posterior means of the $a$ inferred from the data  and the corresponding $y-$axis values, respectively.
\vspace{-1.5mm}
}
\end{figure}

Figure \ref{fig:clusters} (a) shows that the posterior mean of the number of clusters tends to decrease as the discount parameter $a$ increases towards one for the gNBP, but rapidly increases as $a$ increases for the reparameterized gNBP. Figure \ref{fig:clusters} (b) shows that the posterior mean of the ratio of unit-size clusters, as lower bounded by $\max\{a,0\}$ in the prior,  generally increases for both the gNBP and reparameterized gNBP as $a$ increases. Figure \ref{fig:clusters} (c) shows that the posterior mean of the average size of clusters tends to increases as  $a$ increases towards one for the gNBP, but rapidly decreases as $a$ increases for the reparameterized gNBP. Figure \ref{fig:clusters} (d) shows that the posterior mean of the non-unit-size clusters rapidly decrease as  $a$ increases towards one for both the gNBP and reparameterized gNBP. The normalized histograms of the inferred discount parameter $a$ are shown in Figures \ref{fig:clusters} (e) and (f) for the gNBP and reparameterized gNBP, respectively. Note that for the gNBP, %there is a reluctance to rapidly decrease 
the total number of clusters is reluctant to decrease as  $a$ increases towards one, this is because the cluster sizes follow the $\mbox{TNB}(a,p)$ distribution in the prior, which favors a single non-unit-size cluster to be accompanied with  at least $a/(1-a)$ unit-size clusters;  the total number of clusters is also reluctant to increase as the negative $a$ decreases towards $-\infty$, this is because a negative $a$ with a large absolute value would favor the average size of clusters to be large. Although the gNBP and reparameterized gNBP exhibit distinct behaviors on both the number and sizes of clusters, Figures \ref{fig:clusters} (b) and (d) show that they share similar trends on both the ratio and number of non-unit-size clusters as a function of the discount parameter $a$.

\begin{figure}[!tb]
\begin{center}
\includegraphics[width=0.86\columnwidth]{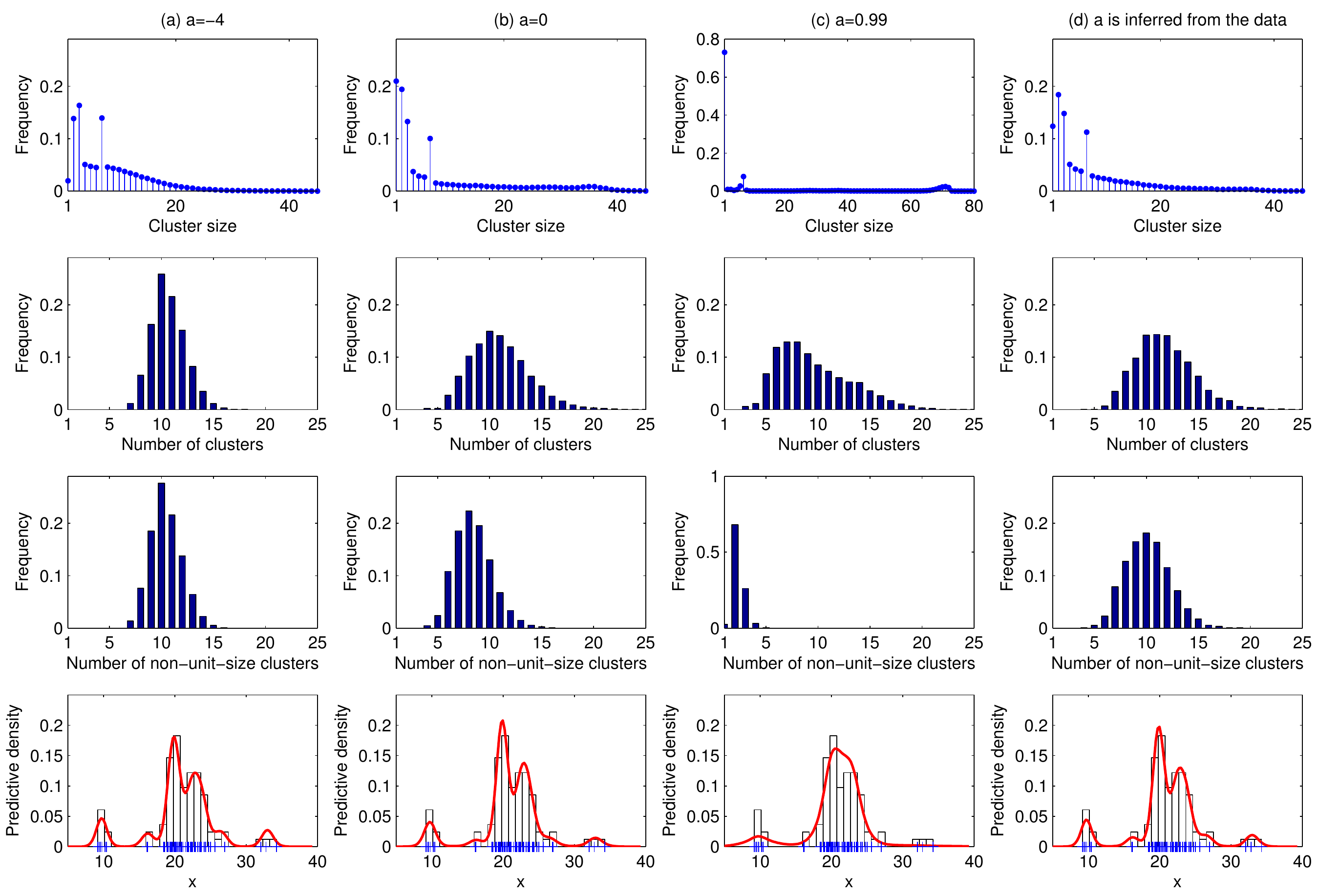}
\end{center}
\vspace{-6.5mm}
\caption{ \label{fig:density} \small
For the generalized negative binomial process Gaussian count-mixture model, rows from top to bottom show the posterior distributions of the sizes of clusters, number of clusters, number of non-unit-size clusters %$\tilde{l}_{(m)}$
and predictive data densities, respectively. Columns from left to right show the results with $a=-4$, $0$ and $0.9$ and  with $a$ inferred from the data, respectively.
\vspace{-1mm}
}
\end{figure}

\begin{figure}[!tb]
\begin{center}
\includegraphics[width=0.86\columnwidth]{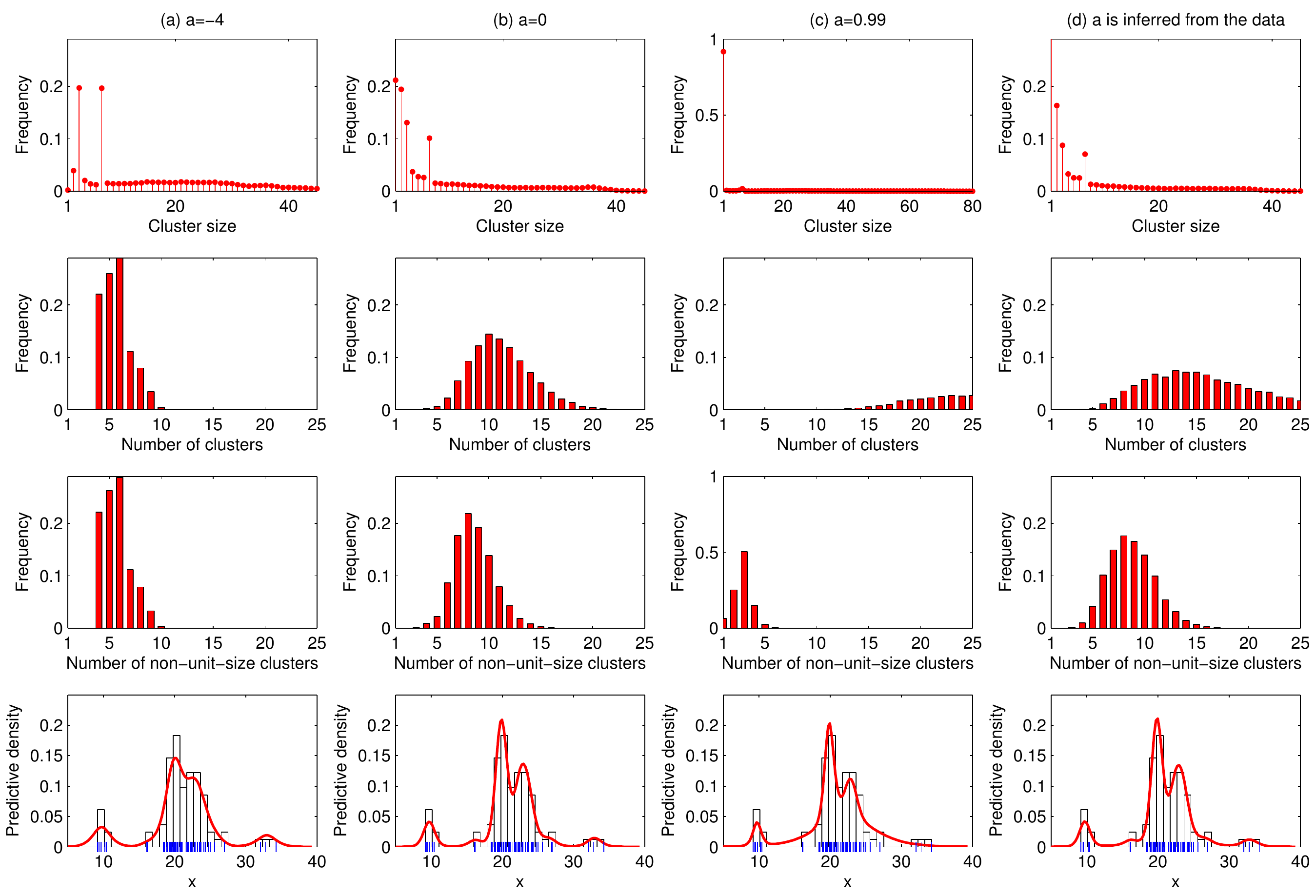}
\end{center}
\vspace{-6.5mm}
\caption{ \label{fig:density_h} \small
For the reparameterized generalized negative binomial process Gaussian count-mixture model, rows from top to bottom show the posterior distributions of the sizes of clusters, number of clusters, number of non-unit-size clusters %$\tilde{l}_{(m)}$
and predictive data densities, respectively. Columns from left to right show the results with $a=-4$, $0$ and $0.9$ and  with $a$ inferred from the data, respectively.
%As a function of the discount parameter $a=-10,0, 0.99$, how the posterior means of the number of non-unit-size clusters $\tilde{l}_{(m)}$ and the ratio of unit-size-clusters $1-\tilde{l}_{(m)}/l{}$ change are shown in (a) and (b), respectively. The histogram of the collected samples for the $a$ that is inferred from the data is shown in (c). The horizontal and vertical positions
%of the $\Box$ sign in both (a) and (b) show the posterior mean of the $a$ inferred from the data and the  posterior mean of the corresponding $y-$axis values, respectively.
\vspace{-1mm}
}
\end{figure}

To better visualize the distinctions between the gNBP and reparameterized gNBP, we compare their posterior cluster-number and cluster-size distributions and predictive densities in Figures \ref{fig:density} and \ref{fig:density_h}. The first row of  Figures \ref{fig:density} shows that for the gNBP, the increase of  $a$ makes the posterior cluster-size distribution not only has a larger probability at $n_k=1$ but also  has heavier tails, encouraging large-size clusters. Whereas the first row of  Figures \ref{fig:density_h} shows that for the reparameterized gNBP, the increase of  $a$ makes the posterior cluster-size distribution has a larger probability at $n_k=1$ and lower probabilities on the tails, discouraging large-size clusters. It is interesting to notice that large posterior probabilities are often assigned to the cluster sizes of $n_k=2$, $n_k=3$ and $n_k=7$. The reason is that there are seven data points around $x=10$, two around $x=16$, two around $x=27$ and three around $x=33$ that are clearly separately from the other ones and hence usually clustered together.
%These observations well match the prior assumptions of the generalized NB process on these values's behaviors.

The second row of Figure \ref{fig:density} shows that for the gNBP, the increase of $a$ drives the high density region of the posterior cluster-number PMF towards left, encouraging fewer clusters, 
%up the posterior probabilities for smaller cluster numbers, 
whereas the second row of Figure \ref{fig:density_h} shows that for the reparameterized gNBP, the increase of $a$ drives the high density region of the posterior cluster-number PMF towards right, 
encouraging more clusters.
%down the posterior probabilities for smaller cluster numbers while drives up posterior probabilities for larger cluster numbers. 
The thirds rows of Figures \ref{fig:density} and \ref{fig:density_h}  show that a discount parameter $a$ close to one would drive down the number of non-unit-size clusters for both the gNBP and reparameterized gNBP. The estimated predictive data densities are shown in the last rows of Figures \ref{fig:density} and \ref{fig:density_h}. 

Figures \ref{fig:density} and  \ref{fig:density_h} clearly show that different discount parameters lead to distinct model behaviors in terms of the distributions of both the number and sizes of clusters. For both the gNBP and reparameterized gNBP, by allowing the learning of $a$, the inferred posteriors have a large support over the values of $a$ in $(-\infty,1)$, covering a wide array of models with distinct cluster structures. % that 

To show that in the posterior,  the EPPFs of both the gNBP $f(z_{1:20}|m,\gamma_0,a,p)$ and  reparameterized gNBP $f(z_{1:20}|m,h_0,a,p)$ are dependent on the sample size $m$, we test both a sample consisting of the $m=20$ galaxies with the lowest relative velocities and a sample consisting of all the $m=82$ galaxies. %change the sample from $x_{1:20}$ to $\xv=x_{1:82}$, 
The parameters are set as $p=0.9$, $\gamma_0=1$, $h_0=1$ and $a=-4$, $-1$, $0$, $0.5$ or $0.9$. It is clear that for both the gNBP and reparameterized gNBP, when $a\neq 0$, with the same parameters $\gamma_0$, $h_0$, $a$ and $p$, the posterior distribution of the number of clusters in $z_{1:20}$ for the sample of $x_{1:20}$ is clearly different from that in $z_{1:20}$ for the sample of $x_{1:82}$. Allowing the partition probability function $P(\Pi_j|m)$ to be dependent on the sample size $m$ is a unique feature of the cluster structure, which is not permitted in partition structures whose EPPFs are subject to the addition rule.

%
%

%\end{align}
%where noninformative priors are imposed on $\mu_0$ and $\varphi_0$ as  $\mu_0\sim\mathcal{N}(0,10^3)$ and $\varphi_0\sim\mbox{Gamma}(10^{-3},1/10^{-3})$.

\begin{figure}[!tb]
\begin{center}
\includegraphics[width=0.9\columnwidth]{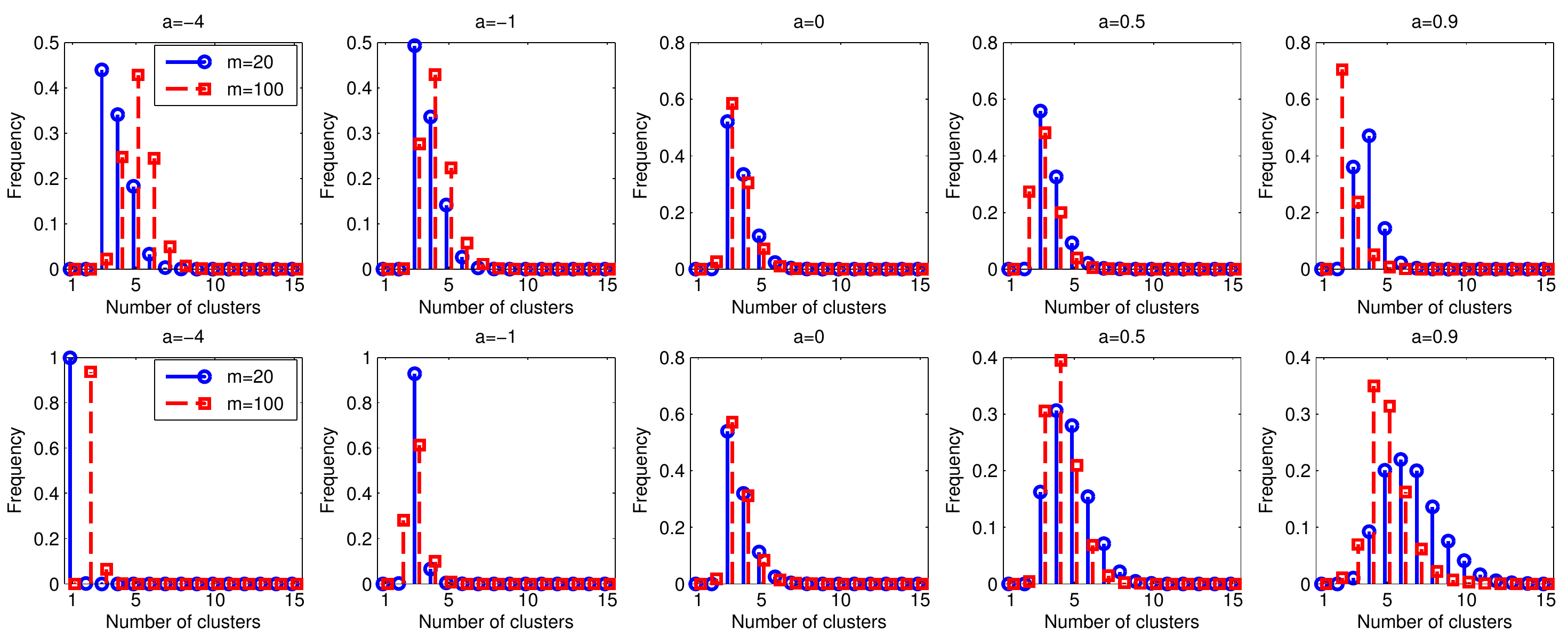}
\end{center}
\vspace{-6mm}
\caption{ \label{fig:sdEPPF_galaxy}  \small
Comparison of the posterior distributions of the number of clusters in $z_{1:20}$ between a sample consisting of the $m=20$ galaxies with the lowest velocities and a sample consisting of all the $m=82$ galaxies, from the Galaxy dataset. The first and second rows show the results for the gNBP and reparameterized gNBP respectively. Columns from left to right are the results for $a=-4$, $-1$, $0$, $0.5$ and $0.9$, respectively. The parameters are set as $\gamma_0=1$, $h_0=1$ and $p=0.9$. 
\vspace{-0mm}
}
\end{figure}

\section{Conclusions}

Every completely random measure with a finite total random mass can be used to construct a count-mixture model, leading to a cluster structure where the cluster-number and cluster-size distributions are determined by the L\'{e}vy measure.  A cluster structure is distinct from a partition structure that the partition probability function of a subset of the sample is allowed to be dependent on the sample size, and the generated exchangeable random partitions are permitted to be inconsistent in distribution as the sample size varies.
The paper presents 
a generalized negative binomial process count-mixture model, 
which generates a cluster structure with a Poisson distributed finite number of clusters and truncated negative binomial distributed cluster sizes. The exchangeable cluster probability function defines a fully factorized marginal likelihood for parameter estimation. The  exchangeable partition probability function, which violates the addition rule when $a<0$ and $0<a<1$, defines a generalized Chinese restaurant process, whose simple prediction rule is used to develop a generalized P\'{o}lya urn sampling scheme.  
Both theoretic analyses and experimental results confirm that the generalized negative binomial process and its reparameterized version can be controlled to exhibit distinct behaviors on both the number and sizes of clusters. 
In both the original and reparameterized versions, the selection of $a \in (-\infty,1)$ is able to effectively reflect one's prior preference on cluster-number and cluster-size distributions, and  the learning of $a$  sidesteps the requirement of model selection,  allowing the posterior to be averaged over a wide array of  models with distinct cluster structures. 
Hierarchical constructions \citep{BNBP_PFA_AISTATS2012,NBPJordan,NBP_NIPS2012} based on the generalized negative binomial process are worth further investigation.

\section*{Acknowledgements} 
The author is grateful to Stephen G. Walker, Lawrence Carin, Fernando A. Quintana and  Peter M\"uller   for their helpful 
comments and suggestions. 

\small
\bibliographystyle{plainnat}
\bibliography{References062013}

\begin{thebibliography}{59}
\providecommand{\natexlab}[1]{#1}
\providecommand{\url}[1]{\texttt{#1}}
\expandafter\ifx\csname urlstyle\endcsname\relax
  \providecommand{\doi}[1]{doi: #1}\else
  \providecommand{\doi}{doi: \begingroup \urlstyle{rm}\Url}\fi

\bibitem[Aalen(1992)]{alen1992modelling}
O.~O. Aalen.
\newblock Modelling heterogeneity in survival analysis by the compound
  {P}oisson distribution.
\newblock \emph{Annals of Applied Probability}, 1992.

\bibitem[Aldous(1983)]{aldous:crp}
D.~Aldous.
\newblock Exchangeability and related topics.
\newblock In \emph{Ecole d'Ete de Probabilities de Saint-Flour XIII}, pages
  1--198. Springer, 1983.

\bibitem[Antoniak(1974)]{DP_Mixture_Antoniak}
C.~E. Antoniak.
\newblock Mixtures of {D}irichlet processes with applications to {B}ayesian
  nonparametric problems.
\newblock \emph{Ann. Statist.}, 1974.

\bibitem[Bar-Lev and Enis(1986)]{bar1986reproducibility}
S.~K. Bar-Lev and P.~Enis.
\newblock Reproducibility and natural exponential families with power variance
  functions.
\newblock \emph{Annals of Statistics}, 1986.

\bibitem[Barrios et~al.(2012)Barrios, Lijoi, Nieto-Barajas, and
  Pruenster]{barrios2012modeling}
E.~Barrios, A.~Lijoi, L.~E. Nieto-Barajas, and I.~Pruenster.
\newblock Modeling with normalized random measure mixture models.
\newblock \emph{Carlo Alberto Notebooks, No. 276}, 2012.

\bibitem[Brix(1999)]{brix1999generalized}
A.~Brix.
\newblock Generalized gamma measures and shot-noise {C}ox processes.
\newblock \emph{Advances in Applied Probability}, 1999.

\bibitem[Broderick et~al.(2013)Broderick, Mackey, Paisley, and
  Jordan]{NBPJordan}
T.~Broderick, L.~Mackey, J.~Paisley, and M.~I. Jordan.
\newblock Combinatorial clustering and the beta negative binomial process.
\newblock \emph{arXiv:1111.1802v5}, 2013.

\bibitem[Charalambides(2005)]{charalambides2005combinatorial}
C.~A Charalambides.
\newblock \emph{Combinatorial methods in discrete distributions}.
\newblock Wiley, 2005.

\bibitem[Engen(1974)]{engen1974species}
S.~Engen.
\newblock On species frequency models.
\newblock \emph{Biometrika}, 1974.

\bibitem[Escobar and West(1995)]{Escobar1995}
M.~D. Escobar and M.~West.
\newblock Bayesian density estimation and inference using mixtures.
\newblock \emph{JASA}, 1995.

\bibitem[Ewens(1972)]{ewens1972sampling}
W.~J. Ewens.
\newblock The sampling theory of selectively neutral alleles.
\newblock \emph{Theoretical Population Biology}, 1972.

\bibitem[Favaro and Teh(2013)]{favaromcmc}
S.~Favaro and Y.~W. Teh.
\newblock {MCMC} for normalized random measure mixture models.
\newblock \emph{to appear in Statistical Science}, 2013.

\bibitem[Ferguson(1973)]{ferguson73}
T.~Ferguson.
\newblock A {B}ayesian analysis of some nonparametric problems.
\newblock \emph{Annals of Statistics}, 1973.

\bibitem[Gerber(1992)]{gerber1992generalized}
H.~U. Gerber.
\newblock From the generalized gamma to the generalized negative binomial
  distribution.
\newblock \emph{Insurance: mathematics and economics}, 1992.

\bibitem[Gnedin and Pitman(2006)]{gnedin2006exchangeable}
A.~Gnedin and J.~Pitman.
\newblock Exchangeable {G}ibbs partitions and {S}tirling triangles.
\newblock \emph{Journal of Mathematical Sciences}, 2006.

\bibitem[Gnedin et~al.(2009)Gnedin, Haulk, and Pitman]{Gnedin_deletion}
A.~Gnedin, C.~Haulk, and J.~Pitman.
\newblock Characterizations of exchangeable partitions and random discrete
  distributions by deletion properties.
\newblock In N.H. Bingham and C.M. Goldie, editors, \emph{Probability and
  Mathematical Genetics: Papers in Honour of Sir John Kingman}. 2009.

\bibitem[Green(2009)]{GreenDP}
P.~Green.
\newblock Colouring and breaking sticks: random distributions and heterogeneous
  clustering.
\newblock In N.H. Bingham and C.M. Goldie, editors, \emph{Probability and
  Mathematical Genetics: Papers in Honour of Sir John Kingman}. 2009.

\bibitem[Griffin(2010)]{griffin2010default}
J.~E. Griffin.
\newblock Default priors for density estimation with mixture models.
\newblock \emph{Bayesian Analysis}, 2010.

\bibitem[Griffin and Walker(2011)]{griffin2011posterior}
J.~E. Griffin and S.~G. Walker.
\newblock Posterior simulation of normalized random measure mixtures.
\newblock \emph{Journal of Computational and Graphical Statistics}, 2011.

\bibitem[Helgen et~al.(2013)Helgen, Pinto, Kays, Helgen, Tsuchiya, Quinn,
  Wilson, and Maldonado]{helgen2013taxonomic}
K.~M. Helgen, C.~M. Pinto, R.~Kays, L.~E. Helgen, M.~T.~N. Tsuchiya, A.~Quinn,
  D.~E. Wilson, and J.~E. Maldonado.
\newblock Taxonomic revision of the olingos ({B}assaricyon), with description
  of a new species, the {O}linguito.
\newblock \emph{ZooKeys}, 2013.

\bibitem[Hjort(1990)]{Hjort}
N.~L. Hjort.
\newblock Nonparametric {B}ayes estimators based on beta processes in models
  for life history data.
\newblock \emph{Ann. Statist.}, 1990.

\bibitem[Hougaard(1986)]{hougaard1986survival}
P.~Hougaard.
\newblock Survival models for heterogeneous populations derived from stable
  distributions.
\newblock \emph{Biometrika}, 1986.

\bibitem[Ishwaran and James(2001)]{ishwaran2001gibbs}
H.~Ishwaran and L.~F. James.
\newblock Gibbs sampling methods for stick-breaking priors.
\newblock \emph{JASA}, 2001.

\bibitem[Ishwaran and Zarepour(2000)]{ishwaran2000markov}
H.~Ishwaran and M.~Zarepour.
\newblock Markov chain {M}onte {C}arlo in approximate {D}irichlet and beta
  two-parameter process hierarchical models.
\newblock \emph{Biometrika}, 2000.

\bibitem[James et~al.(2009)James, Lijoi, and Pr{\"u}nster]{james2009posterior}
L.~F. James, A.~Lijoi, and I.~Pr{\"u}nster.
\newblock Posterior analysis for normalized random measures with independent
  increments.
\newblock \emph{Scandinavian Journal of Statistics}, 2009.

\bibitem[Johnson et~al.(2005)Johnson, Kemp, and Kotz]{johnson2005univariate}
N.~L. Johnson, A.~W. Kemp, and S.~Kotz.
\newblock \emph{Univariate Discrete Distributions}.
\newblock John Wiley \& Sons, 2005.

\bibitem[J{\o}rgensen(1997)]{jorgensen1997theory}
B.~J{\o}rgensen.
\newblock \emph{The Theory of Dispersion Models}.
\newblock London: Chapman \& Hall, 1997.

\bibitem[Kingman(1967)]{Kingman}
J.~F.~C. Kingman.
\newblock Completely random measures.
\newblock \emph{Pacific Journal of Mathematics}, 1967.

\bibitem[Kingman(1978{\natexlab{a}})]{kingman1978random}
J.~F.~C. Kingman.
\newblock Random partitions in population genetics.
\newblock \emph{Proceedings of the Royal Society of London. A.},
  1978{\natexlab{a}}.

\bibitem[Kingman(1978{\natexlab{b}})]{kingman1978representation}
J.~F.~C. Kingman.
\newblock The representation of partition structures.
\newblock \emph{Journal of the London Mathematical Society},
  1978{\natexlab{b}}.

\bibitem[Kingman(1993)]{PoissonP}
J.~F.~C. Kingman.
\newblock \emph{Poisson Processes}.
\newblock Oxford University Press, 1993.

\bibitem[Lee et~al.(2013)Lee, Quintana, M{\"u}ller, and Trippa]{lee2013}
J.~Lee, F.~A. Quintana, P.~M{\"u}ller, and L.~Trippa.
\newblock Defining predictive probability functions for species sampling
  models.
\newblock \emph{Statistical Science}, 2013.

\bibitem[Lijoi and Pr{\"u}nster(2010)]{BeyondDP}
A.~Lijoi and I.~Pr{\"u}nster.
\newblock Models beyond the {D}irichlet process.
\newblock In N.~L. Hjort, C.~Holmes, P.~M{\"u}ller, and S.~G. Walker, editors,
  \emph{Bayesian nonparametrics}. Cambridge University Press, 2010.

\bibitem[Lijoi et~al.(2005)Lijoi, Mena, and
  Pr{\"u}nster]{lijoi2005inverseGaussian}
A.~Lijoi, R.~H. Mena, and I.~Pr{\"u}nster.
\newblock Hierarchical mixture modeling with normalized inverse-{G}aussian
  priors.
\newblock \emph{Journal of the American Statistical Association}, 2005.

\bibitem[Lijoi et~al.(2007)Lijoi, Mena, and Pr{\"u}nster]{lijoi2007controlling}
A.~Lijoi, R.~H. Mena, and I.~Pr{\"u}nster.
\newblock Controlling the reinforcement in {B}ayesian non-parametric mixture
  models.
\newblock \emph{Journal of the Royal Statistical Society: Series B}, 2007.

\bibitem[MacEachern and M{\"u}ller(1998)]{MullerDP}
S.~N. MacEachern and P.~M{\"u}ller.
\newblock Estimating mixture of {D}irichlet process models.
\newblock \emph{Journal of Computational and Graphical Statistics}, 1998.

\bibitem[M{\"u}ller and Mitra(2013)]{Muller2013}
P.~M{\"u}ller and R.~Mitra.
\newblock Bayesian nonparametric inference -- why and how.
\newblock \emph{Bayesian Analysis}, 2013.

\bibitem[M{\"u}ller and Quintana(2004)]{muller2004nonparametric}
P.~M{\"u}ller and F.~A. Quintana.
\newblock Nonparametric {B}ayesian data analysis.
\newblock \emph{Statistical Science}, 2004.

\bibitem[Neal(2000)]{neal2000markov}
R.~M. Neal.
\newblock Markov chain sampling methods for {D}irichlet process mixture models.
\newblock \emph{Journal of computational and graphical statistics}, 2000.

\bibitem[Perman et~al.(1992)Perman, Pitman, and Yor]{perman1992size}
M.~Perman, J.~Pitman, and M.~Yor.
\newblock Size-biased sampling of poisson point processes and excursions.
\newblock \emph{Probability Theory and Related Fields}, 1992.

\bibitem[Pitman(1995)]{pitman1995exchangeable}
J.~Pitman.
\newblock Exchangeable and partially exchangeable random partitions.
\newblock \emph{Probability Theory and Related Fields}, 1995.

\bibitem[Pitman(1996)]{Pitman96somedevelopments}
J.~Pitman.
\newblock Some developments of the {B}lackwell-{M}acqueen urn scheme.
\newblock In \emph{Statistics, Probability and Game Theory; Papers in honor of
  David Blackwell}, 1996.

\bibitem[Pitman(2003)]{pitman2003poisson}
J.~Pitman.
\newblock Poisson-{K}ingman partitions.
\newblock \emph{Lecture Notes-Monograph Series}, pages 1--34, 2003.

\bibitem[Pitman(2006)]{csp}
J.~Pitman.
\newblock \emph{Combinatorial stochastic processes}.
\newblock Lecture Notes in Mathematics. Springer-Verlag, 2006.

\bibitem[Pitman and Yor(1997)]{pitman1997two}
J.~Pitman and M.~Yor.
\newblock The two-parameter {P}oisson-{D}irichlet distribution derived from a
  stable subordinator.
\newblock \emph{The Annals of Probability}, 1997.

\bibitem[Quenouille(1949)]{LogPoisNB}
M.~H. Quenouille.
\newblock A relation between the logarithmic, {P}oisson, and negative binomial
  series.
\newblock \emph{Biometrics}, 1949.

\bibitem[Regazzini et~al.(2003)Regazzini, Lijoi, and
  Pr{\"u}nster]{regazzini2003distributional}
E.~Regazzini, A.~Lijoi, and I.~Pr{\"u}nster.
\newblock Distributional results for means of normalized random measures with
  independent increments.
\newblock \emph{Annals of Statistics}, 2003.

\bibitem[Ritter and Tanner(1992)]{griddygibbs}
C.~Ritter and M.~A. Tanner.
\newblock Facilitating the {G}ibbs sampler: the {G}ibbs stopper and the
  griddy-{G}ibbs sampler.
\newblock \emph{Journal of the American Statistical Association}, 1992.

\bibitem[Roeder(1990)]{roeder1990density}
K.~Roeder.
\newblock Density estimation with confidence sets exemplified by superclusters
  and voids in the galaxies.
\newblock \emph{JASA}, 1990.

\bibitem[Thibaux and Jordan(2007)]{JordanBP}
R.~Thibaux and M.~I. Jordan.
\newblock {Hierarchical beta processes and the Indian buffet process}.
\newblock In \emph{AISTATS}, 2007.

\bibitem[Tweedie(1984)]{tweedie1984index}
M.~C.~K. Tweedie.
\newblock An index which distinguishes between some important exponential
  families.
\newblock In \emph{Statistics: Applications and New Directions: Proc. Indian
  Statistical Institute Golden Jubilee International Conference}, pages
  579--604, 1984.

\bibitem[Willmot(1988)]{willmot1988remark}
G.~E. Willmot.
\newblock A remark on the poisson-pascal and some other contagious
  distributions.
\newblock \emph{Statistics \& probability letters}, 1988.

\bibitem[Wolpert and Ickstadt(1998)]{Wolpert98poisson/gammarandom}
R.~L. Wolpert and K.~Ickstadt.
\newblock Poisson/gamma random field models for spatial statistics.
\newblock \emph{Biometrika}, 1998.

\bibitem[Wolpert et~al.(2011)Wolpert, Clyde, and Tu]{Wolp:Clyd:Tu:2011}
R.~L. Wolpert, M.~A. Clyde, and C.~Tu.
\newblock Stochastic expansions using continuous dictionaries: {L\'e}vy
  {A}daptive {R}egression {K}ernels.
\newblock \emph{Annals of Statistics}, 2011.

\bibitem[Zhou and Carin(2012)]{NBP_NIPS2012}
M.~Zhou and L.~Carin.
\newblock Augment-and-conquer negative binomial processes.
\newblock In \emph{NIPS}, 2012.

\bibitem[Zhou and Carin(2013)]{NBP2012}
M.~Zhou and L.~Carin.
\newblock Negative binomial process count and mixture modeling.
\newblock \emph{arXiv:1209.3442, accepted for publication in IEEE Trans.
  Pattern Analysis and Machine Intelligence}, 2013.

\bibitem[Zhou et~al.(2011)Zhou, Yang, Sapiro, Dunson, and
  Carin]{dHBP_AISTATS2011}
M.~Zhou, H.~Yang, G.~Sapiro, D.~Dunson, and L.~Carin.
\newblock Dependent hierarchical beta process for image interpolation and
  denoising.
\newblock In \emph{AISTATS}, 2011.

\bibitem[Zhou et~al.(2012{\natexlab{a}})Zhou, Chen, Paisley, Ren, Li, Xing,
  Dunson, Sapiro, and Carin]{BPFA_TIP2012}
M.~Zhou, H.~Chen, J.~Paisley, L.~Ren, L.~Li, Z.~Xing, D.~Dunson, G.~Sapiro, and
  L.~Carin.
\newblock Nonparametric {B}ayesian dictionary learning for analysis of noisy
  and incomplete images.
\newblock \emph{IEEE Trans. Image Processing}, 2012{\natexlab{a}}.

\bibitem[Zhou et~al.(2012{\natexlab{b}})Zhou, Hannah, Dunson, and
  Carin]{BNBP_PFA_AISTATS2012}
M.~Zhou, L.~Hannah, D.~Dunson, and L.~Carin.
\newblock Beta-negative binomial process and {P}oisson factor analysis.
\newblock In \emph{AISTATS}, 2012{\natexlab{b}}.

\end{thebibliography}

\end{document}